\documentclass[acmsmall]{acmart}

\AtBeginDocument{%
  }

\setcopyright{cc}
\setcctype{by-sa}
\acmJournal{PACMHCI}
\acmYear{2025} \acmVolume{9} \acmNumber{7} 
\acmArticle{CSCW508}
\acmMonth{11} \acmPrice{}\acmDOI{10.1145/3757689}

\usepackage{CJK}

\begin{document}

\title[Understanding Digital Gifting Through Messengers Across Cultures]{Understanding Digital Gifting Through Messengers Across Cultures: A Comparative Study of University Students in South Korea, China, and Japan}

\author{YeEun Lee}
\affiliation{%
 \institution{Seoul National University}
 \city{Seoul}
 \country{South Korea}}
\email{jigtm1107@snu.ac.kr}

\author{Dakyeom Ahn}
\affiliation{%
  \institution{Seoul National University}
  \city{Seoul}
  \country{South Korea}}
\email{adklys@snu.ac.kr}

\author{JungYu Kwon}
\affiliation{%
  \institution{Seoul National University}
  \city{Seoul}
  \country{South Korea}}
\email{jessey2000@snu.ac.kr}

\author{SeungJi Lee}
\affiliation{%
  \institution{Seoul National University}
  \city{Seoul}
  \country{South Korea}}
\email{leesjee2000@snu.ac.kr}

\author{Hajin Lim}
\authornote{Corresponding author.}

\affiliation{%
  \institution{Seoul National University}
  \city{Seoul}
  \country{South Korea}}
\email{hajin@snu.ac.kr}

\renewcommand{\shortauthors}{YeEun Lee, Dakyeom Ahn, JungYu Kwon, SeungJi Lee, and Hajin Lim}

\keywords{Online gifting, digital gifting, instant messenger, mobile gifting platforms, Kakaotalk, Line, Wechat, hongbao, cross-cultural studies, culturally sensitive design, social reciprocity}

\begin{abstract}
Digital gift-giving has become a key means of maintaining social relationships, but most existing research has focused on gifting within global e-commerce or social media platforms. The emergence of messenger-based gifting in East Asia, where Korea, Japan, and China each have distinct and deeply rooted gifting traditions, remains underexplored. This study examines how in-app gifting services on the most widely used messaging platforms in South Korea (KakaoTalk), Japan (LINE), and China (WeChat) reflect and reshape culturally embedded gifting practices. Through semi-structured interviews with 26 university students, we found that KakaoTalk facilitates frequent, informal exchanges aligned with Korea's emphasis on broad social ties; LINE supports selective and carefully presented gifts, reflecting Japanese norms of formality and sincerity; and WeChat's Hongbao feature enables playful, communal monetary exchanges largely detached from traditional, obligation-driven gifting. Drawing on these findings, we propose the Channel-Oriented Gifting Cycle model, which extends classical gift-exchange theory by showing that the choice of gifting platform is not merely logistical but a culturally meaningful part of the gifting process. We conclude with design implications for culturally sensitive digital gifting services.
\end{abstract}

\ccsdesc[500]{Human-centered computing~Empirical studies in collaborative and social computing}

\maketitle

\begin{CJK}{UTF8}{mj}

\section{Introduction}
Gift-giving serves as a fundamental social practice for expressing emotions, signaling intimacy, and strengthening relationships \cite{sherry1983gift, belk1979gift}. Traditionally, gift exchanges have played a central role in sustaining social bonds by conveying cultural meanings and fulfilling various relational objectives \cite{belk1979gift, schwartz1967social}. In the digital age, these practices have evolved markedly, shifting from the in-person exchange of physical goods to digitally mediated forms of expression \cite{sandel1998money, yang2011virtual, hampton2011social}.

Digital gifting---the exchange of gifts primarily facilitated through digital platforms \cite{kwon2022exploring, kwon2017s}---has emerged as a prominent social practice where the entire process from selection to reception occurs electronically. While research has extensively examined digital gift-giving within global e-commerce and social media contexts \cite{amazon2024, close2010beyond, kizilcec2018social}, less is known about how these practices unfold within instant messaging platforms, where gifting is often conversational, spontaneous, and shaped by platform affordances. In particular, little attention has been paid to how digital gifting is embedded in culturally specific norms of reciprocity, intimacy, and social obligation. This gap invites closer examination of how digital infrastructures mediate and reconfigure long-standing cultural traditions of gift exchange. 

In particular, East Asian messaging platforms offer a rich context for addressing this gap. Unlike Western counterparts that primarily serve more globalized or general-purpose communication, platforms such as KakaoTalk (South Korea), LINE (Japan), and WeChat (China) are deeply localized and nationally dominant, with user penetration rates of 93\%, 81\%, and 81\% respectively as of 2024 \cite{choi2024kakaotalk, straight2024line, kiran2024wechat}). Beyond messaging, these platforms have developed elaborate in-app gifting features that are closely aligned with local cultural norms and rituals. They function as nationally embedded social infrastructures, shaping how users enact reciprocity, celebrate occasions, and express care in everyday interaction.

Over the past decade, gifting functionalities on East Asian messaging platforms have evolved into integral components of everyday social interactions. For example, KakaoTalk introduced its ``Gift'' feature in 2010, allowing users to send digital gift vouchers directly through the chat interface \cite{kakaogift2010}. LINE followed with ``LINE Gift'' service in 2015 \cite{line2021gift}, and WeChat launched its widely adopted ``Red Envelope (\begin{CJK}{UTF8}{gbsn}红包\end{CJK}, Hongbao)'' feature in 2014 \cite{wechat2024Hongbao2}. These features, commonly used for birthdays, holidays, and expressions of gratitude \cite{kwon2022exploring}, represent a meaningful extension of traditional gifting practices within technologically mediated communication. Interestingly, although these platforms build on the similar infrastructure of instant messaging, their gifting features differ markedly in design, functionality, and usage patterns, which likely reflect the influence of local cultural values and social norms.

Despite geographical proximity and shared Confucian influences, South Korea, Japan, and China each exhibit distinct gift-giving customs shaped by nuanced cultural traditions. These distinctive cultural landscapes offer a valuable context to examine not only how digital gifting platforms adapt to culturally specific norms but also how they actively reshape traditional practices within each social context. Yet, existing literature has largely overlooked how messenger-based digital gifting intersects with these evolving cultural dynamics in contemporary East Asian societies.

To address this gap, we conducted in-depth interviews with 26 university students in their twenties from Korea, Japan, and China, each with over a year of experience using their country's digital gifting services through messengers. We deliberately focused on this demographic, as young adults are among the most active users of digital media and are navigating a key transitional life stage \cite{junco2008introduction}. During this period, they are forming, expanding, and reshaping social networks across personal and professional domains \cite{burke2016moving}, making them particularly inclined to developing new relationship management strategies through digital platforms.

Specifically, our study addresses three key research questions:

\begin{itemize}
\item\textbf{RQ1.} What are the distinctive usage patterns of messenger-based gifting services among university students in Korea, Japan, and China?
\item\textbf{RQ2.} How do cultural values and norms shape the similarities and differences in digital gift-giving practices across these three countries?
\item\textbf{RQ3.} In what ways are messenger-based gifting services reshaping traditional gift-giving practices within each cultural context?
\end{itemize}

Our findings reveal how digital gift-giving practices among young adults both reflect and redefine culturally embedded social norms. In South Korea, digital gifting significantly expands strategies for personalization and relationship-building, representing a digital extension of traditional practices. In Japan, by contrast, digital gifting has evolved into more streamlined, standardized interactions that emphasize presentation and etiquette in line with cultural expectations. In China, the \textit{Hongbao} feature has facilitated new forms of group-based interaction, though face-to-face exchanges rooted in \textit{guanxi} remain highly influential.

By bridging the gap between traditional theories of gift-giving with contemporary digital practices, this study contributes to a deeper theoretical understanding of how cultural contexts shape and are shaped by digital social technologies. 
To articulate this process, we propose the \textbf{Channel-Oriented Gifting Cycle} model, an extension of classical gift-exchange theory that foregrounds the culturally meaningful role of gifting platforms in shaping how gifts are selected, interpreted, and received. Whereas earlier models treated the gifting channel as a secondary logistical step, our framework shows that the choice of gifting platform fundamentally mediates emotional intent, social appropriateness, and relational expectations within specific cultural contexts.

In addition to its theoretical contributions, this study provides design implications for creating culturally sensitive digital gifting platforms. By examining gifting practices in South Korea, Japan, and China, we highlight how culturally rooted norms can guide the development of gifting features that support emotional expression, relational nuance, and cultural adaptability.


\section{Related Work}

\subsection{Gift-Giving as a Means of Relationship Maintenance}
The act of giving gifts has long been recognized as a key mechanism for maintaining interpersonal relationships \cite{mauss1967gift}. Gift-giving serves not only to express the giver's affection and celebration but also to reflect and reinforce the social ties between the giver and the recipient \cite{belk1979gift, belk1993gift}. As both symbolic and material exchanges, gifts encapsulate shared memories and emotions, serving a range of relational purposes \cite{berking1999sociology}. Central to these social exchanges is the principle of reciprocity \cite{gray2014paying}, which fosters mutual support and indebtedness, thereby sustaining social bonds \cite{levi-strauss1969kinship, malinowski1978argonauts, mauss1954gift}.

The theoretical foundation for understanding gift-giving as a socially embedded practice originates from Mauss's seminal work on gifting \cite{mauss1954gift}. Mauss conceptualized gift exchange as more than merely economic; rather, he viewed it as a central mechanism through which social bonds are created and sustained. In his framework, gift-giving entails a cyclical process of obligation: to give, to receive, and to reciprocate. This foundational idea has been widely expanded and refined by other scholars.

A particularly influential extension is offered by Sherry \cite{sherry1983gift}, who delineated gift-giving into three stages: \textbf{Gestation}, \textbf{Prestation}, and \textbf{Reformulation}. In the \textbf{Gestation} stage, the giver carefully considers not only the recipient's characteristics but also the symbolic meaning of the gift and its situational context. The \textbf{Prestation} stage involves the ritualized transfer of the gift, where emotional and social meanings are visibly expressed through practices such as gift-wrapping and ceremonial presentation. As the final stage, \textbf{Reformulation} involves the recipient's interpretation of the gift and its integration into the relationship, potentially reshaping social dynamics. Sherry emphasized these stages as part of an ongoing cycle that both reinforces and evolves relational ties. 

The rise of digital technologies has significantly transformed the context and mechanics of gift exchange. To address these changes, Kwon et al. \cite{kwon2017s} proposed a model tailored to digital environments, arguing that digital gifts differ notably from physical ones in terms of experiential characteristics and interactive affordances. Their model outlines a five-stage sequence: selection, delivery, reception, usage, and reflection. Notably, they observe that digital gifting tends to involve weaker emotional investment during the early stages---particularly selection and delivery---and that its relational significance tends to dissipate quickly during reflection.

Despite these theoretical advancements, existing frameworks still present notable limitations. Current models pay insufficient attention to how platform-specific delivery mechanisms shape the Gestation stage by influencing what is gifted, to whom, and under what social cues. Additionally, cultural variations in gift exchange remain relatively underexplored. While Sherry's model provides valuable analytical tools for understanding the affective and symbolic dimensions of gifting, it falls short of addressing how cultural influences distinctly shape digital gifting across diverse contexts. These gaps call for a more nuanced understanding of digital gift-giving that considers both platform design and sociocultural variability.


\subsection{Cultural differences of Gift Giving among Korea, China, and Japan} 
Though the culture of gift exchange is a universal phenomenon, it holds particularly significant meaning in collective societies such as those in East Asia, which are often contrasted with Western cultures centered around individualism \cite{lee2012korean, lonner1980culture, hui1986individualism}. In Korea, Japan, and China, gift-giving has evolved beyond a simple exchange of goods into a crucial ritual that helps balance social relationships, affirms individuals' roles within the community, and reinforces alignment with societal norms \cite{li2022differentiating, rupp2003gift, kim2024politics}. As a result, gift exchanges often function as a form of  ``social contract'', where the trust and obligations they create play a central role in sustaining long-term relationships.

While there are common elements of gift-giving culture across these three countries---such as emphasizing the giver's attentiveness and the importance of maintaining harmony through thoughtful reciprocity \cite{vanbaal1975reciprocity, luo1997guanxi}---each country has also developed its own distinct customs and gifting practices. 

Beginning with South Korea, gift-giving is deeply influenced by Confucian values \cite{lonner1980culture, lee1991cross}. A central tenet of Confucianism is the concept of ``face'', which emphasizes maintaining dignity, respect, and appropriate behavior relative to one's social status \cite{ho1976concept, lee1988cross}. This notion strongly shapes Korean gifting practices. Reciprocation is not only expected but also seen as a way of preserving face and honoring social obligations. As a result, Korean gift-giving often reflects both a psychological need to repay kindness and a culturally embedded concern for sustaining harmony and hierarchical balance \cite{kim2006giftgiving}.

In the Chinese context, a distinctive cultural concept is \textit{guanxi}  \cite{cadsby2008trust}, which refers to a system of interpersonal relationships rooted in mutual trust, obligation, and long-term reciprocity. Guanxi functions as an informal social infrastructure that often supplements or substitutes for formal institutions \cite{xin1996guanxi, walder1988communist}.  Within this networked structure, gift-giving plays a vital role in initiating and maintaining social ties \cite{cadsby2008trust}. Traditional material gifts---such as food, tea, or furniture---are commonly exchanged during major holidays like the Spring Festival and Mid-Autumn Festival, reinforcing social bonds \cite{chen2013chinese}.  Additionally, monetary gifts in red envelopes, known as Hongbao (\begin{CJK}{UTF8}{gbsn}红包\end{CJK}, red packets), are widely given at weddings, birthdays, and festivals. More than a financial gesture, hongbao is a ritualized practice that symbolizes relational continuity, mutual trust, and social cohesion \cite{saxena2024digitalising, bian2019guanxi}.

In Japan, gift-giving is marked by a strong emphasis on formality, sincerity, and aesthetic presentation. Gifts are valued not merely for their material content, but as a medium for expressing respect, gratitude, and emotional care \cite{rupp2003gift}. The cultural practice of tsutsumi (\begin{CJK}{UTF8}{min}包み\end{CJK}, gift-wrapping) is particularly significant in this context, with traditional wrapping cloths such as furoshiki (\begin{CJK}{UTF8}{min}風呂敷\end{CJK}) used to convey the giver's thoughtfulness and attentiveness \cite{suarez2020understanding}. The meaning of a gift also varies depending on the nature of the relationship. In reciprocal relationships, recipients often feel a social obligation to return the gesture. In hierarchical relationships, by contrast, reciprocation is less expected, as the act of giving may serve to acknowledge one's indebtedness or respect for status differences \cite{beff1984zotou, suzuki1988gift}. Situational norms further shape gifting expectations. For instance, in the custom of okaeshi (\begin{CJK}{UTF8}{min}お返し\end{CJK}, return gifts), the returned gift is typically valued at about half the original, a convention that expresses gratitude while carefully avoiding implications of social distance or imbalance \cite{rupp2003gift, ulkuniemi2022comparison}.

Despite these subtle yet significant cultural differences, Korea, Japan, and China have frequently been grouped together under Western theoretical frameworks---particularly the individualism-collectivism dichotomy---due to their shared historical and cultural intersections \cite{lonner1980culture, hui1988measurement, triandis1990cross}. This has contributed to an assumption of cultural homogeneity among these nations \cite{park1998comparison}, leading many studies to examine gift-giving practices in each country independently rather than through direct comparison. Departing from this tendency, our study explicitly investigates how each country's unique relational values shape the design and use of messenger-based gift-giving interfaces. 

\subsection{Gift Giving Technology}

While traditional gifting has predominantly relied on face-to-face interactions, the rise of digital technologies has significantly transformed gifting practices \cite{sandel1998money}. Digital gifting now encompasses a wide range of formats,  including gift cards, virtual items exchanged through social networks, and digital goods shared within online communities. 

Research on digital gifting within HCI began gaining traction in the early 2000s, initially focusing on how mobile communication technologies enabled teenagers to exchange digital items that fostered relational intimacy \cite{taylor2002age}. Later studies, however, revealed that explicitly labeling digital items as ``gifts'' could diminish their perceived sincerity relative to physical gifts \cite{kwon2017s}. In response, scholars have proposed hybrid forms of gifting---integrating digital elements with physical objects---as a way to enhance relational meaning and complement traditional gift-giving rituals \cite{spence2023more, koleva2020designing}.

Much of the existing literature has focused on global commerce and social media platforms such as Facebook and Amazon, examining how social visibility, product selection, and delivery mechanisms shape gifting experiences \cite{watts2015social, gunasti2018200}. However, messenger-based gifting platforms operate quite differently. They embed gift selection, personalization, and delivery within ongoing conversational contexts, where relational meanings are shaped not only by the gift itself but by its timing, context, and interactional cues \cite{lee2020makes}. These structural differences highlight the need to examine messenger-based gifting as a culturally embedded social practice rather than merely a transactional feature.

Within East Asia---particularly South Korea, China, and Japan---digital gifting features are actively integrated into dominant messenger platforms such as KakaoTalk, WeChat, and LINE. While WeChat's hongbao (digital red envelope) feature has received substantial scholarly attention in the Chinese context \cite{wu2017money, yang2011virtual, liu2023wechat}, research on messenger-based gifting in South Korea and Japan remains limited. Studies in Korea have primarily examined contexts such as mobile vouchers and reciprocity motivations on KakaoTalk \cite{Lee2013DigitalGifts, Lee2017Gifticon, Kwon2021MobileReturnGifts}, while academic work on LINE-based gifting practices in Japan is notably sparse.

Our study aims to address these gaps by comparatively analyzing digital gifting practices across KakaoTalk, WeChat, and LINE. We explore how gifting behaviors on these platforms are shaped by culturally specific norms and relational expectations. In doing so, we highlight that digital gifting technologies are not culturally neutral or universally experienced; rather, they function as socially embedded systems, shaped by and reflective of the local traditions, values, and interpersonal practices of the societies in which they operate.

\section{Method}
We conducted semi-structured interviews with 26 university students in their 20s from Korea, Japan, and China to explore their usage patterns, emotions, and relationship management related to messenger-based gifting services. Participants were active users of the dominant messaging platform in their country (KakaoTalk, LINE, or WeChat), each of which reflects local social norms. The interviews yielded qualitative insights into how digital gift-giving practices are embedded in everyday communication and shaped by platform-specific features. 

We focused on university students in their twenties due to their high proficiency and active engagement with digital platforms, including online gifting services \cite{niu2022machiavellianism}. As a generation who experienced the transition from traditional, face-to-face gifting practices during childhood \cite{taylor2002age} to the widespread adoption of digital gifting tools \cite{kizilcec2018social}, these young adults were uniquely positioned to reflect on changes in gifting practices. Additionally, their familiarity with digital platforms also reduced potential technological barriers, allowing us to focus more directly on sociocultural variation. This age group is also navigating a transitional social phase, where relationships expand beyond close personal ties to include academic, professional, and institutional networks \cite{burke2016moving}. While not unique to students, the university setting offers a focused social environment well-suited for examining how digital gifting aligns with evolving relational expectations and culturally embedded norms \cite{de2023social, yang2011virtual}.

\subsection{Participant Recruitment}
We recruited participants in their twenties from South Korea (2 males, 9 females), Japan (4 males, 3 females), and China (2 males, 6 females), all of whom had used the dominant messenger-based gifting platforms in their country---KakaoTalk, LINE, and WeChat---for at least one year. All participants were fluent in either Korean or English.
In total, we recruited 11 participants from Korea, 7 from Japan, and 8 from China. Participants were recruited through university student communities, social media platforms, and international exchange student networks. For Japanese and Chinese participants, we additionally employed snowball sampling to expand the participant pool.

\subsection{Interview Procedures} 
We conducted semi-structured interviews, offering participants the choice of in-person or remote sessions based on their preference. Interviews lasted approximately one hour and were conducted by a two-person research team: one researcher led the conversation while the other took notes. 

Our interview questions were organized around three research goals. To explore RQ1 (usage patterns), we asked about the frequency, context, and typical situations for using gifting features. For RQ2 (cultural meanings), we inquired into gift selection, appropriateness, and whether choices reflected personal or cultural values. For RQ3 (shifts in tradition), participants compared digital and offline gifting and reflected on evolving social expectations around gift-giving. As participants responded, they were encouraged to recall and share recent or particularly memorable digital gifting experiences. When they felt comfortable, they also shared screenshots of relevant interactions. 

To ensure ethical research practices, all participants received a detailed consent form outlining the study's purpose, procedures, privacy protections, and their rights. Each participant received a gift voucher worth approximately 20,000 KRW (about \$14) immediately after the interview. The study protocol was approved by the Institutional Review Board (IRB) of Seoul National University.

\subsection{Data Analysis}
We employed thematic analysis \cite{patton2014qualitative} to examine both the cultural dimensions of gifting practices and how participants across the three countries use messenger-based gifting services for relationship management. Interviews were recorded and transcribed using ClovaNote, and the transcripts were securely stored on a cloud drive accessible only to the research team. Screenshots shared by participants during the interviews were anonymized and embedded into the corresponding transcripts.
To protect participant privacy, each individual was assigned a unique Subject ID. Analysis was conducted in either Korean or English, depending on the participant's language preference. Korean transcripts were carefully translated into English by bilingual members of the research team during the writing phase to ensure consistency and preserve linguistic nuance.

After removing identifiable information and correcting transcription errors, five authors collaboratively conducted the thematic analysis to identify recurring themes within and across countries. Each author independently reviewed the transcripts multiple times and initially categorized themes into three areas: traditional gifting culture, culturally shaped practices within the digital service, and newly emerging gifting behaviors unique to the digital environment.

Themes identified across researchers were then synthesized into higher-level thematic clusters. To ensure analytic rigor, all authors participated in multiple rounds of collaborative discussion to review, refine, and validate the emerging themes. We used Miro as a shared workspace to support collaborative coding and visualization throughout the analysis.

\section{Findings}

We present our findings by first outlining the traditional offline gift-giving practices in China, Japan, and Korea, and then examining how these established cultural norms have been reconfigured in the digital realm through mobile messenger-based gifting services. Drawing on Sherry's three-phase framework \cite{sherry1983gift}---Gestation, Prestation, and Reformulation---we highlight the cultural continuity and distinct transformations occurring as gifting practices transition from offline to digital contexts.

\subsection{Gift-Giving Traditions in the Three East Asian Countries}

Through interviews with participants from the three countries, we reconfirmed a shared understanding that gift-giving involves a sense of social obligation and requires careful selection to express goodwill. However, the ways in which these exchanges have been carried out vary considerably according to each country's distinct cultural values and social expectations. By examining historical gift-giving practices, we illustrate how cultural contexts have traditionally shaped relational behaviors and set the foundation for adopting and using messenger-based gifting services.

\subsubsection{Gift-Giving Traditions in China}
Chinese participants explained that gift-giving is strongly influenced by the concept of \textit{guanxi}, a system of interpersonal relationships built on mutual obligations and trust. Within this cultural framework, gifts were regarded as one of the most powerful tools for strengthening social ties and expressing goodwill. In certain situations, offering a gift was even considered a prerequisite for gaining recognition or approval from the other party. For example, C3 stated,  \textit{``In some situations, you have to give a gift for the other person to acknowledge you.''}

In this context, materiality played a central role in signaling sincerity. Participants described physical gifts as essential to demonstrating sincerity and fulfilling social obligations. Non-material forms of giving, such as services, were frequently viewed as inadequate or insincere. C3 noted, \textit{``There's a strong belief that a gift must be something tangible. If it's not material, it can seem insincere, like you didn't really put in any effort.''}

Interestingly, such emphasis on gift-giving in public or formal \textit{guanxi}-based relationships appeared to discourage gift exchanges in more private contexts. Because gifts in formal settings often carry an explicit purpose, offering a gift outside of these contexts could be interpreted as an attempt to gain favor or manipulate the relationship. As a result, participants reported that personal gift exchanges were typically reserved for very close relationships, where the giver's intentions were less likely to be misunderstood: \textit{``People avoid giving gifts unless they're close, as it may seem like they expect something in return due to guanxi.''} (C5)

In this way, gifts in China function not merely as expressions of affection but as strategic instruments for building trust and gaining recognition. Relationship formation and maintenance are shaped by social norms that emphasize the exchange of tangible resources, clearly distinguishing transactional public relationships from more intimate, carefully managed private ties.

\subsubsection{Gift-Giving Traditions in Japan}
Japanese participants described gift-giving as a deeply embedded cultural practice, shaped by strong social expectations and ritualized norms intended to maintain relational harmony. Although practices such as \textit{omiyage}---souvenirs given after traveling---suggest frequent gift exchanges even among casual acquaintances, participants noted these exchanges have become so routine and ritualistic that they no longer meaningfully impact relationship dynamics. As one participant explained, \textit{``Giving omiyage after a trip is such a given that it doesn't really feel like a traditional gift anymore.''} (J2)

Compared to these routinized exchanges, deliberate gift-giving intended to build or deepen relationships---especially with less familiar acquaintances---was described as relatively uncommon. Participants emphasized concerns about potential misinterpretation, noting that selecting a gift often required careful consideration to avoid unintended implications. One participant, for example,  described cosmetics as particularly sensitive gifts, explaining, \textit{``With cosmetics, for example, it could come across as implying dissatisfaction with someone's makeup, so you really have to be careful.''} (J4)

As a result, participants indicated a preference for neutral, generic items with minimal potential for misinterpretation. Several participants expressed concern that gift choices perceived as too personal or specific might inadvertently lead to discomfort or misunderstanding, leading them to reserve such exchanges for carefully maintained relationships.

Another recurring theme across Japanese participants was the importance of \textit{wrapping} in Japanese gift culture. Participants frequently mentioned that the presentation of the gift could sometimes be perceived as equally significant as the gift itself. One participant reflected, \textit{``Receiving an unwrapped gift feels a bit indifferent. It might even make me worried. I've never received a gift that wasn't wrapped. Even with snacks, they're always in some kind of wrapper or bag. Wrapping is an important part of our culture.''} (J1) 

These accounts suggest that interpersonal connections in Japan tend to be managed through cautious and carefully regulated gift exchanges, particularly in private settings. The emphasis on care, discretion, and presentation reflects a culturally specific approach to relationship management, in which the act of giving is shaped as much by form and subtlety as by content.

\subsubsection{Gift-Giving Traditions in Korea}
Korean participants commonly described gift-giving as a meaningful practice for managing and sustaining social relationships. These practices occurred not only during major occasions such as birthdays, but also in a range of everyday contexts involving casual exchanges. Several participants noted that gift-giving was common even among acquaintances or within less intimate relationships. In such cases, giving and receiving material gifts were described as a familiar and socially expected routine.

Participants further emphasized that gifts were not merely expressions of goodwill but purposeful acts used to initiate or reinforce a broad range of social connections. Many described an implicit understanding that material gifts carried an expectation of reciprocity, creating ongoing cycles of exchange that helped maintain relational balance.  As one participant explained, \textit{``Because gift-giving has become a kind of `formality,' there's this reciprocal sense that if you receive something material, you have to give something back---so people end up exchanging gifts repeatedly.''} (K5)

This strong expectation of reciprocity was especially pronounced in less intimate relationships. In such cases, gift exchanges were often perceived as transactional, accompanied by a clear social pressure to reciprocate promptly. One participant noted, \textit{``If we're not really close, giving gifts just feels like a transaction. The more awkward the relationship, the stronger the urge to quickly return the favor when I receive something.''} (K6)

These accounts suggest that gift exchanges in Korea may serve as a key mechanism for sustaining broader and more loosely connected social networks. Further, it highlights how regular, reciprocal exchanges help maintain relational continuity---even in casual or distant relationships---by signaling attentiveness and reinforcing mutual obligation.

In sum, while traditional gift-giving practices in all three countries serve as important mechanisms for forming and maintaining social relationships, our findings reveal notable cross-cultural differences shaped by each country's distinctive values and relational norms. Chinese participants tended to view gifts as pragmatic tools, strategically exchanged within clearly defined relational boundaries. Japanese participants characterized their gift-giving as cautious and restrained, shaped by concerns about potential misunderstandings and social discomfort. Korean participants emphasized ongoing, reciprocal exchanges that appeared to support the maintenance of broader and more loosely connected social networks.

\subsection{Mobile Messenger-Based Digital Gift-Giving in the Three Countries}
In this section, we examine how traditional gift-giving norms are enacted and adapted within mobile messenger-based digital gifting contexts, highlighting both continuities and emergent practices shaped by the affordances of digital platforms.

To structure our analysis, we draw on Sherry's three-phase framework of gift-giving---\textbf{Gestation, Prestation, and Reformulation} \cite{sherry1983gift}. In this framework, \textbf{Gestation} refers to the initial planning stage, where the giver considers the recipient, occasion, and intent of the gift. Next, \textbf{Prestation} involves the act of giving itself, particularly the moment the gift is delivered. Finally, \textbf{Reformulation} addresses how the recipient receives and interprets the gift, and how the exchange might influence or reshape the future of the relationship.

We applied this framework to analyze how mobile messenger-based digital gifting practices in China, Japan, and Korea align with---or diverge from---the three phases of gift-giving as they are typically enacted in offline contexts. In doing so, we also examine differences in the extent of adoption and localization of digital gifting services across the three countries. 


\subsubsection{WeChat Gifting - China}

\textit{\textbf{Gestation Phase}}.
Participants described gift-giving in Chinese culture primarily as a deliberate, strategic practice shaped by the relational logic of \textit{guanxi}. During the Gestation phase---when deciding on a gift and planning how to deliver it---participants emphasized the importance of choosing items that could explicitly signal respect, recognition, or obligation within formal or public relationships. In this initial planning stage, digital gifts were often dismissed as inappropriate or ineffective. Many participants expressed concern that digital channels obscured key relational gestures and introduced risks by leaving traceable records, raising issues of privacy and social appropriateness within the \textit{guanxi} framework. As participant C5 explained, \textit{``Gift-giving is heavily influenced by guanxi. Most gifts are exchanged offline, so online gifting feels unfamiliar.''} Another participant explicitly emphasized privacy concerns: \textit{``Guanxi doesn't happen online. This is due to the concern that online exchanges, which leave a visible record, can feel culturally awkward or less appropriate.'} (C5) These accounts suggest that, during the planning stage, digital gifting via mobile messengers was rarely seen as suitable for fulfilling the relational obligations embedded in traditional \textit{guanxi}-driven exchanges.

\begin{figure}[b] 
  \includegraphics[width=0.55\textwidth]{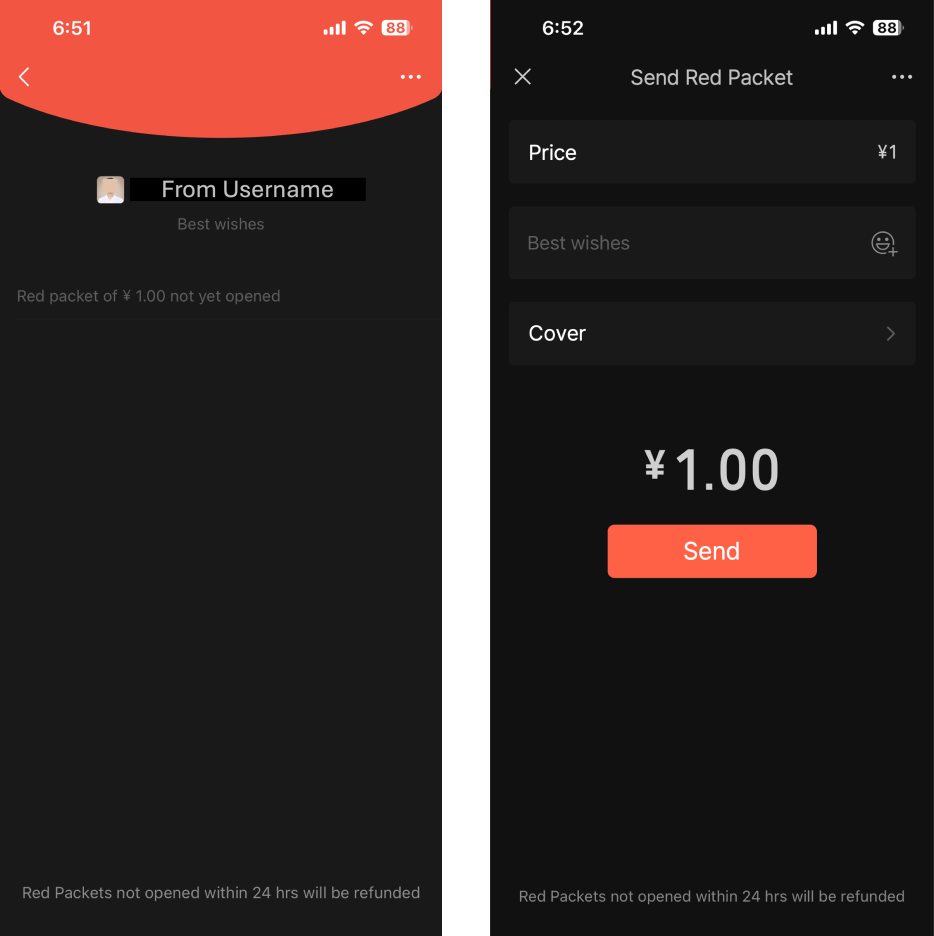}
  \captionsetup{justification=centering}
  \caption{Receiving and Sending Hongbao (Red Packet) in WeChat}
  \Description{A screenshot from WeChat showing both receiving and sending Hongbao. In the receiving view, the user can see the sender's username, message, and the Hongbao amount. In the sending view, the user can specify the amount, message, and cover design for the Hongbao.}
  \label{fig:hongbao}
\end{figure}

\textit{\textbf{Prestation Phase}}.
Participants indicated that traditional gifting in China typically involved direct, tangible exchanges, which contrasted sharply with the more limited material-based digital gifting observed on WeChat. Instead, the Prestation phase---the act of gift delivery---was primarily expressed through the widespread use of the \textit{Hongbao} feature (See Fig.\ref{fig:hongbao}) as a convenient, culturally meaningful, and socially acceptable mode of digital gifting. 
Many noted its ease and informality, highlighting how it enabled lightweight expressions of care without the logistical burden of physical gifts. One participant noted, \textit{``Hongbao makes it easier to show your heart than a physical gift. You can even embed meaning in the numbers.''} (C6) This use of numerically coded values---such as amounts referencing lucky numbers or inside jokes---was seen as a form of symbolic expression unique to \textit{Hongbao} gifting, offering a playful yet emotionally resonant alternative to conventional gifts.

\textit{\textbf{Reformulation Phase}}.
In terms of how gifts influenced relationships after delivery, participants reported that \textit{Hongbao} exchanges fostered new, informal patterns of interaction, distinct from the formal relational obligations typically associated with traditional gifting. Rather than reinforcing hierarchical ties or reciprocal expectations, \textit{Hongbao} was described as facilitating playful, collective engagements, particularly in group chats. These digital exchanges encouraged social bonding through lightweight, gamified interactions. As C5 explained: \textit{``There's a game-like culture where the person who gets the largest amount starts the next round.''} (C5) 


Overall, Sherry's three-phase framework reveals how digital gifting via WeChat diverges meaningfully from traditional offline gifting practices. In the \textbf{Gestation} phase, participants largely rejected digital gifting due to concerns about appropriateness, traceability, and the relational authenticity central to \textit{guanxi}. In the \textbf{Prestation} phase, rather than replicating offline gift-giving, WeChat introduced the culturally specific practice of \textit{Hongbao}, enabling symbolic and numerical expression of sentiments. Finally, in the \textbf{Reformulation} phase, \textit{Hongbao} shifted the social function of gifting from obligation-based reciprocity to collective play and informal bonding. This illustrates how digital gifting in China selectively adapts traditional norms while also creating new relational possibilities beyond established \textit{guanxi}-based expectations.

\subsubsection{LINE Gifting - Japan}
\textit{\textbf{Gestation Phase}}.
During the initial planning stage of digital gifting, participants reported carefully considering factors such as the recipient, the occasion, and the nature of the relationship. Digital gifting via LINE was generally seen as appropriate for casual, low-stakes situations---such as minor occasions, simple expressions of gratitude, or brief apologies---contexts requiring minimal emotional investment or formality.  As one participant noted clearly: \textit{``I use it when it's not a friend I see often, or when I just want to say sorry.''} (J1)

Conversely, participants explicitly reported avoiding digital gifting when planning for more formal or emotionally significant relationships. They associated traditional gifting strongly with sincerity, emotional depth, and careful presentation---qualities participants felt were not adequately conveyed through digital channels. Particularly, several participants highlighted the cultural significance of wrapping and physical presentation, citing these as reasons for favoring physical gifts. J1 explained: \textit{``I always use gift-wrapping services at stores. If a gift isn't wrapped, it feels a bit careless. Wrapping is considered a natural part of giving a gift.''}

\begin{figure}[b] 
  \includegraphics[width=0.35\textwidth]{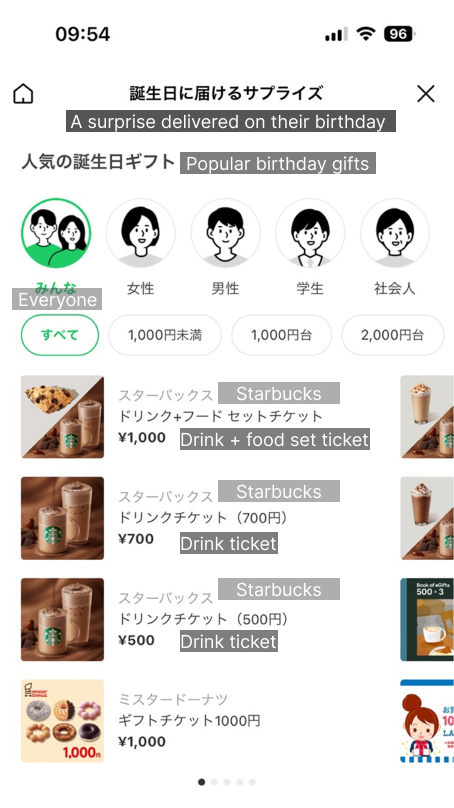}
  \captionsetup{justification=centering}
  \caption{Starbucks as the primary recommended item in LINE's birthday gift user interface}
  \Description{A screenshot from Line. This image shows LINE's birthday gift recommendation interface, where Starbucks drink tickets dominate the suggested items. }
  \label{fig:starbucks}
\end{figure}

Consequently, when digital gifting was considered acceptable, participants reported selecting simple, generic items, such as Starbucks coffee coupons, to minimize social risk and potential misinterpretation. One participant succinctly summarized this approach: \textit{``Whether we're close or not, I just send Starbucks.''} (J2) (See Fig.\ref{fig:starbucks})

Alternatively, when offline gifting was not feasible, participants deliberately chose LINE's virtual birthday card feature over material digital gifts. A carefully written digital card was seen as a more sincere and culturally appropriate gesture, aligning with expectations around emotional thoughtfulness and relational sensitivity. 

\textit{\textbf{Prestation Phase}}.
Participants described the act of delivering digital gifts via LINE (the Prestation phase) as significantly more streamlined than traditional physical exchanges.
While most opted for generic items, several described attempts to integrate traditional etiquette into digital exchanges by utilizing LINE's customizable digital message cards. These cards were seen as a way to compensate for the absence of physical wrapping, helping to convey care and thoughtfulness.  J3 detailed their approach: \textit{``When I send a gift, I put in a lot of effort on the card. Even if it's a hassle, I still make one 90\% of the time. A nice illustration or long message feels like it replaces the wrapping.''} 

Nevertheless, participants expressed persistent dissatisfaction with digital gifting, citing its lack of physical wrapping and the visibility of the gift's monetary value. J2 participant emphasized discomfort around this transparency: \textit{``With online gifts, the price is right there. But when you wrap a gift at a store, they usually hide the price.''} (J2) These reflections underscore ongoing concerns about whether digital gifting could meet cultural expectations for careful presentation, emotional subtlety, and aesthetic discretion.

\textit{\textbf{Reformulation Phase}}.
Among Japanese participants, digital gifts were generally perceived as one-time emotional gestures rather than tools for sustaining ongoing relational ties. Because the selection process was streamlined---often involving simple, low-risk items like coffee coupons---participants expressed relatively low expectations for reciprocation or extended exchange. 

Applying Sherry's three-phase framework to LINE-based digital gifting highlights both continuities and departures from traditional gift-giving norms in Japan. In the \textbf{Gestation} phase, participants typically limited digital gifting to casual or informal contexts, selecting neutral and symbolic items to minimize relational risks. During the \textbf{Prestation} phase, some elements of physical gifting---such as wrapping and message personalization---were adapted through LINE's digital card features, though concerns about price visibility and lack of careful presentation persisted.  In the \textbf{Reformulation} phase, digital gifts were interpreted as modest affirmations of existing relationships rather than invitations to deepen or formalize relational commitments. Together, these findings suggest that participants adopted digital gifting practices selectively, aligning them with broader Japanese cultural values that prioritize sincerity, subtlety, and cautious relationship management.

\subsubsection{KakaoTalk Gifting - Korea}
\textit{\textbf{Gestation Phase}}.
Participants described the initial gift-planning phase as deeply embedded within everyday communication practices on KakaoTalk. In this Gestation phase---where individuals consider the recipient, occasion, and appropriateness of the gift---digital gifting was widely seen as acceptable and convenient across diverse relational contexts, including casual acquaintances as well as emotionally significant or hierarchical relationships. Notably, participants explicitly indicated that KakaoTalk gifting was frequently employed not only when face-to-face interactions were challenging, but even among individuals who met regularly in person, including in formal relationships such as with supervisors or professors. K3 summarized this widespread adoption: \textit{``More than 90\% of gifts are online now. It's now a part of the culture.''} (K3)

Participants further explained that messenger-based gifting had expanded beyond formal events and special occasions to become a routine social behavior. Everyday interactions---such as offering comfort, encouragement, or spontaneous gratitude during chats---often trigger gift exchanges on KakaoTalk. Illustrating this habitual practice, K2 shared:
\textit{``When a friend seems to be having a hard time, or when the mood feels heavy during a chat, I send coffee or dessert coupons. I was touched when I received such gifts, so I started doing the same.''} (K2)

Taken together, these accounts suggest that in the Gestation phase, KakaoTalk gifting has transformed from an occasional or event-based act into a normalized component of daily relational maintenance.

\begin{figure}[t] 
  \includegraphics[width=0.9\textwidth]{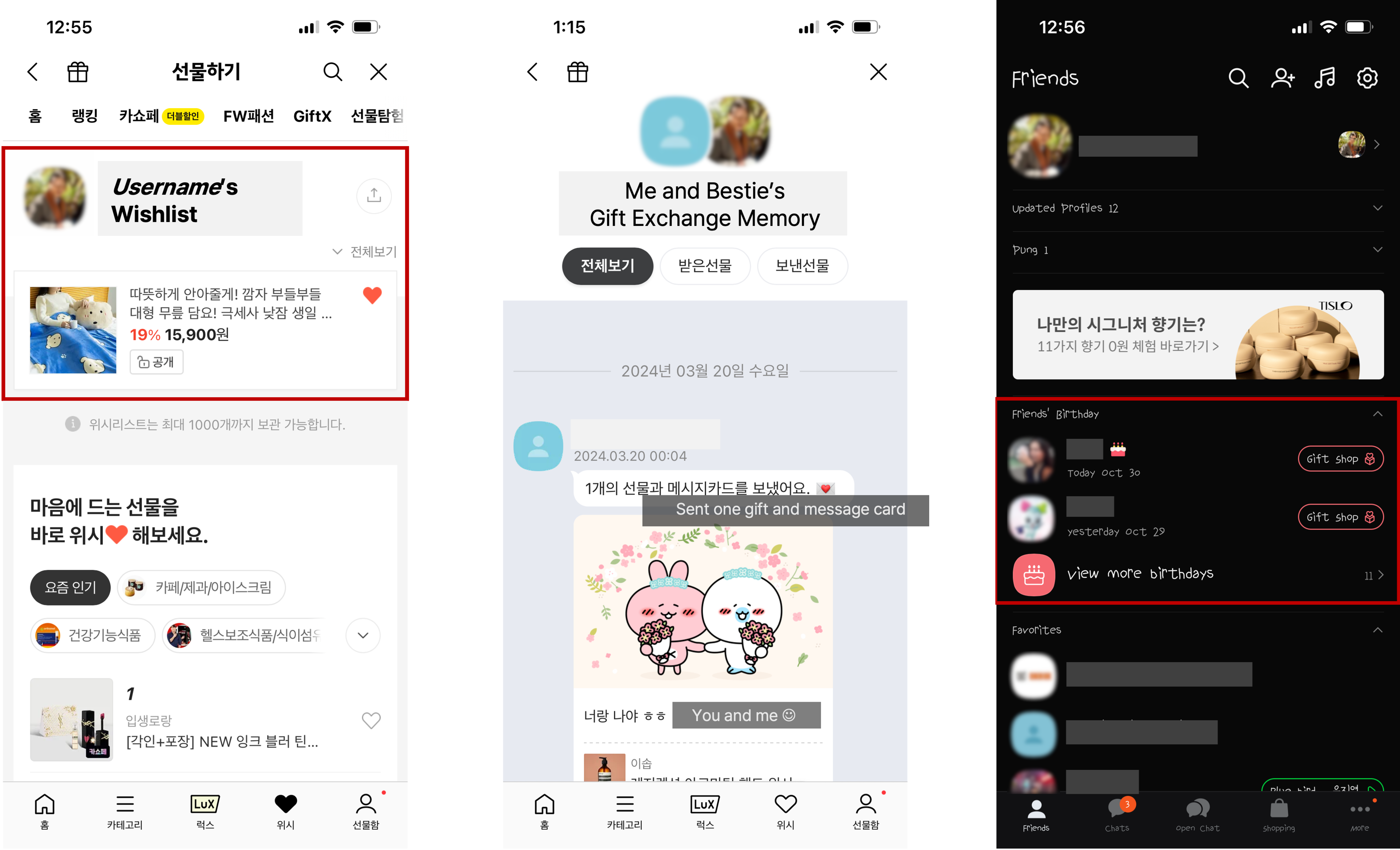}
  \captionsetup{justification=centering}
  \caption{Key gifting features in KakaoTalk: Wishlist, Gift History, and Birthday List (left to right).}
    \Description{
    The first screenshot shows a user's Wishlist in KakaoTalk, including item names, prices, and options to share or set visibility. The second screenshot displays the Gift History, showing a timeline of sent and received gifts along with message cards. The third screenshot presents the Birthday List of friends, each with a button that allows the user to send a gift directly.}
  \label{fig:kakaotalk}
\end{figure}

\textit{\textbf{Prestation Phase}}.
The Prestation phase---centered on the actual selection and sending of digital gifts---revealed participants' careful attention to conveying sincerity, despite the inherent convenience of online gifting.  While KakaoTalk offers automated gift suggestions and integrated wishlists to simplify the gift selection process (See Fig.\ref{fig:kakaotalk}, left), participants frequently described deliberately avoiding such options. Instead, many emphasized making intentional, personalized choices to signal thoughtfulness and relational care. For example,  K9 emphasized careful personalization during digital gift selection:
\textit{``I try to choose something that perfectly fits the person, based on what they've said or a situation I remember.''} (K9) These accounts suggest that participants actively balanced the efficiency of digital gifting with deliberate acts of customization, aiming to preserve sincerity and authenticity throughout the exchange.

\textit{\textbf{Reformulation Phase}}.
In the Reformulation phase, which focused on how digital gift exchanges reshaped relational dynamics after receipt, participants repeatedly highlighted intensified expectations of reciprocity, which were reinforced by KakaoTalk's interface. Participants frequently mentioned that the platform's explicit affordances, such as birthday notifications, gift-history tracking, and labeling senders visibly as ``gift this friend,'' significantly amplified implicit social obligations around gifting (See Fig.\ref{fig:kakaotalk}, middle and right). As K5 explained, these affordances contributed to a growing sense of reciprocal obligation: \textit{``Gifting has become a kind of formal etiquette, so when I receive something material, I automatically feel I need to give something back''}. Further illustrating this reciprocity pressure, K6 noted explicit considerations around matching the value of past gifts: \textit{``I remember the price range of past gifts so I can match it next time''} (K6).

As a result, many participants expressed discomfort regarding these heightened reciprocity norms, particularly in less intimate relationships. Some reported feeling pressure to continually reciprocate gifts, leading to relational fatigue. As one participant shared, the visibility of birthdays and gifting prompts on KakaoTalk created unwanted pressure to engage in performative reciprocity, even with distant acquaintances: \textit{``I didn't want to feel like I had to celebrate the birthdays of people I don't really care about, so I turned off birthday notifications.''} (K8)

Additionally, participants noted that digital gifting was increasingly reshaping how emotional support was expressed in everyday interactions. Many felt that verbal messages alone, such as offering comfort through chat, were beginning to feel insufficient without an accompanying material gesture. K4 described this emergent pressure: 
\textit{``It feels like emotions have to be expressed through gifts now. Words alone feel insufficient.''} (K4)

Overall, drawing on Sherry's three-phase framework clarifies how digital gifting via KakaoTalk distinctly transformed traditional Korean gifting norms. In the \textbf{Gestation} phase, digital gifting became broadly normalized and integrated into everyday interactions across diverse relational contexts, extending across both casual and formal relational contexts. In the \textbf{Prestation} phase, participants actively balanced the convenience of digital platforms with intentional, personalized expressions of relational care. Finally, in the \textbf{Reformulation} phase,  platform-driven reciprocity features intensified cultural expectations for balanced, visible exchanges, heightening social and emotional pressure. Together, these dynamics suggest that KakaoTalk gifting significantly reshaped relational practices by formalizing reciprocity expectations, amplifying performative obligations, and introducing new emotional labor into technology-mediated social life.

\section{Discussions}

This study investigated how traditional gift-giving practices in Korea, Japan, and China are enacted and adapted through mobile messenger-based digital gifting platforms. In this section, we describe three key implications of our findings. First, we extend Sherry's gift-exchange framework by incorporating the culturally contingent choice of delivery channel as a structurally significant element within the gifting cycle.  Second, we examine how digital gifting technologies become integrated into everyday practices, highlighting culturally specific continuities and emergent behaviors. Finally, we propose design implications to support the design of culturally sensitive and socially meaningful digital gifting experiences.

\subsection{The Channel-Oriented Gifting Cycle} 
To explore how digital technologies and cultural norms reshape the structure of gift exchange, we revisit Sherry's classical gift-giving cycle and propose an extended framework informed by our cross-cultural findings. 

Sherry's original model outlines three phases through which gift exchange unfolds: \textbf{Gestation, Prestation, and Reformulation} \cite{sherry1983gift}. Traditionally, the Gestation phase involves the giver's internal deliberation about what gift to give, shaped by factors such as the recipient, relationship type, and occasion \cite{weinberger2025relational}. Conversely, logistical decisions, such as how and through which channel the gift will be delivered (e.g., in-person or by mail), have typically been located within the Prestation phase \cite{kwon2017s}. That is, the delivery method has conventionally been treated as a practical concern, secondary or peripheral to the relationally symbolic act of gift selection.

However, our findings suggest that digital gifting challenges and expands this traditional view by prompting individuals to explicitly consider the delivery channel much earlier, directly within the Gestation phase. Across Korea, Japan, and China, participants consistently described digital gifting platforms not merely as logistical tools but as structurally influential elements that shaped the initial framing of gift exchange. 


In Korea, the pervasive acceptance and integration of KakaoTalk Gifting illustrated how digital platforms were considered from the very start of the gifting process, often preceding the explicit choice of the recipient or occasion. In Japan, by contrast, concerns about emotional authenticity and appropriateness led to more conditional engagement with digital gifting platforms, limiting the types of gifts deemed acceptable. In China, deeply rooted \textit{guanxi}-based relational norms largely excluded digital channels from traditional gift planning. Instead, digital gifting was confined to discrete practices such as \textit{Hongbao}, which operated under a parallel but culturally distinct logic.

\begin{figure}[t] 
  \includegraphics[width=\textwidth]{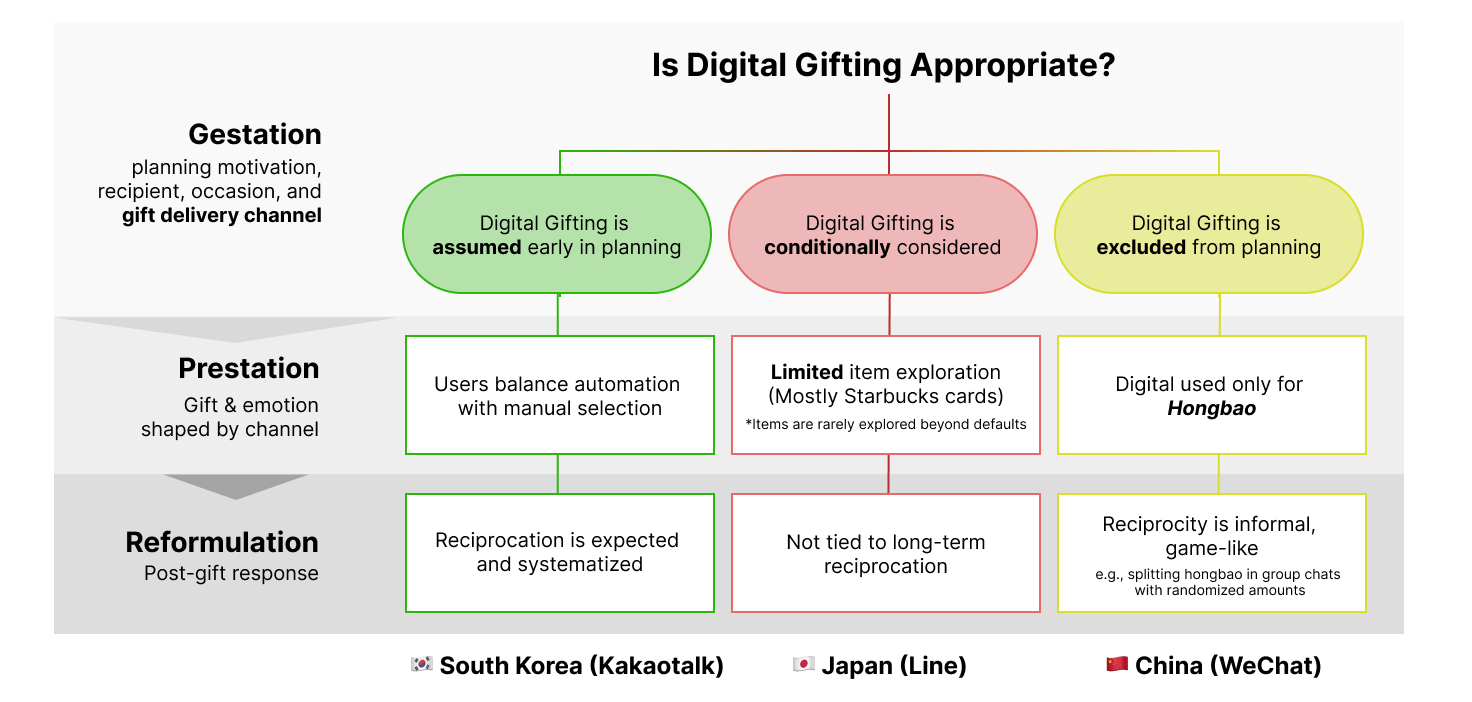}
  \captionsetup{justification=centering}
  \caption{The Channel-Oriented Gifting Cycle applied to findings from South Korea, Japan, and China}
  \Description{A framework illustrates how the choice of gift channel is embedded from the Gestation phase and influences the nature of gifting across cultures. }
  \label{fig:model}
\end{figure}

To account for this conceptual and structural shift, we propose the Channel-Oriented Gifting model (see Fig.\ref{fig:model}), which expands Sherry's original conception of the Gestation phase. Whereas Sherry's model emphasizes internal deliberation about the gift itself, our revised framework explicitly incorporates the culturally contingent selection of the delivery channel as an integral part of the Gestation phase. This choice is shaped not only by relational context but also by platform-specific affordances---such as item variety, personalization features, and available modes of emotional expression. These affordances are not merely functional; they actively reflect and shape the giver's deliberations, influencing both what is chosen and how emotional intent is communicated.

By foregrounding the role of delivery channels in gift planning, the Channel-Oriented Gifting model offers both a theoretical lens for understanding how digital infrastructures structure gift exchange and a comparative framework for examining how culturally situated values are negotiated through platform-mediated practices.

\subsection{Cultural Adoption of Digital Gifting Technologies}
\begin{figure}[t] 
  \includegraphics[width=\textwidth]{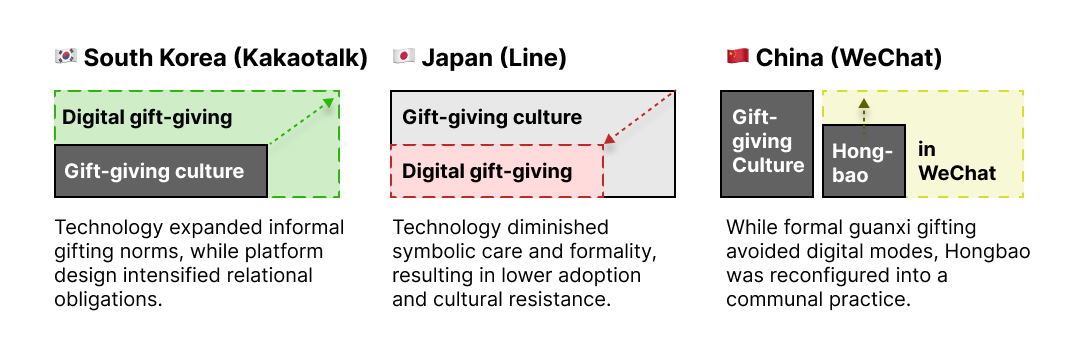}
  \captionsetup{justification=centering}
  \caption{Cultural integration of digital gifting in Korea, Japan, and China. While South Korea saw digital gifting expand informal norms and intensify relational expectations through platform features, Japan's emphasis on symbolic care and formality led to cautious, low-stakes adoption. In China, traditional \textit{guanxi}-based gifting remained largely offline, while Hongbao was reimagined through WeChat as a distinct, communal digital practice.}
  \Description{The figure compares how digital gift-giving aligns with each country's gifting culture. Korea embraced informal gifting, Japan showed resistance, and China adapted Hongbao into a communal digital practice.}
  \label{fig:countries}
\end{figure}

Our findings further provide rich cultural accounts of how digital gifting technologies become integrated into everyday practices, highlighting both continuities with traditional norms and emergent patterns. Drawing on interviews with participants from Korea, Japan, and China, our results show that similar technological features---such as mobile messenger-based gift delivery---are appropriated and imbued with culturally specific ways. Rather than uniformly adapting to technological affordances, participants described digital gifting as amplifying, diminishing, or reconfiguring cultural norms surrounding reciprocity, emotional expression, and social obligation. (See Fig.\ref{fig:countries})

These dynamics unfolded in notably different ways across the three countries.

In South Korea, mobile messenger-based gifting amplified existing cultural practices of maintaining broad, loosely connected social networks. Platform features such as birthday reminders and instant gift exchanges increased the spontaneity and frequency of gifting, embedding gifting more deeply within routine social interactions. At the same time, the ease and visibility of digital gifting heightened feelings of social obligation and relational burden, reinforcing cultural expectations that emotional expressions be accompanied by material gestures.

In Japan, digital gifting was adopted selectively and with caution, reflecting longstanding cultural expectations around sincerity, formality, and careful presentation. Because digital platforms typically lack traditional elements such as gift-wrapping rituals and face-to-face delivery, participants confined their use to casual or low-stakes contexts. As a result, gift selections often defaulted to neutral, convenience-oriented items such as coffee coupons. This selective adoption reflected a deliberate effort to preserve cultural expectations of sincerity by reserving digital exchanges for relationships where emotional risk was low and symbolic meaning was minimal.

In China, digital gifting practices centered on \textit{Hongbao} emerged as a culturally resonant yet distinct mode of exchange. While participants continued to value traditional gifting within guanxi-based relationships, they described digital Hongbao as a separate practice, primarily associated with playful occasions and group interactions. Although the symbolic value of monetary exchange remained intact, digital platforms facilitated rapid, large-scale participation through features such as mass distributions and gamified group draws. Rather than replacing traditional material gifting, digital \textit{Hongbao} was understood as a parallel form of interaction--one more closely aligned with playfulness and collective bonding than with formal reciprocity or hierarchical obligation \cite{augustin2024digitalising}.

Taken together, these cross-cultural differences demonstrate that digital gifting technologies are not merely neutral tools; rather, they become culturally meaningful through continuous processes of social negotiation, resulting in diverse sociotechnical arrangements and contextually grounded practices.This reaffirms that technology adoption is not deterministic, but culturally situated---actively shaped and reinterpreted through ongoing social practices \cite{orlikowski1992duality, 
Akrich1992de, silverstone1996design}.

\subsection{Design Implications for Culturally Sensitive Online Gifting}

Our comparative analysis of mobile messenger-based gifting practices in Korea, Japan, and China underscores that digital gifting is not merely transactional but deeply embedded within culturally specific social practices. Building on these findings, we propose design considerations to support culturally sensitive digital gifting experiences. These implications can be particularly relevant for global e-commerce and messaging platforms, where diverse cultural expectations around emotional expression, reciprocity, and symbolic meaning need to be thoughtfully navigated.

\subsubsection{Preserve the Emotional Core of Gifting}
Across all cultures, gift-giving is not merely a material transaction---it is a social and emotional act grounded in care, intention, and symbolic meaning. Participants in all three countries highlighted that the value of a digital gift derives from the emotional effort it conveys. Korean users described investing in personalized gifts to express sincerity. Japanese participants used carefully crafted messages to compensate for the absence of traditional rituals like wrapping or in-person delivery. Also, in China, participants noted that emotional meaning was often conveyed through the specific amount of money in a \textit{Hongbao}.

These findings underscore a central design challenge: digital platforms, optimized for efficiency, often strip gifting of its emotional depth. To preserve the core function of gifting, platforms must offer features that enable expressive and intentional interaction. This could include customizable messages, voice, or video embeds that help convey context and care. Without such affordances, digital gifts risk losing the very meaning that makes them socially valuable across cultures.

\subsubsection{Honor Ritual and Symbolism as Core to Gifting}
Gifting is a ritualized practice across cultures, grounded not just in the exchange of items but in the symbolic forms and presentation styles that make gifts socially and emotionally meaningful. These rituals---whether through wrapping, timing, or symbolic gestures---signal intention, respect, and cultural belonging. When these elements are absent or misaligned in digital spaces, the social legitimacy of a gift may be diminished.

Our findings reflect this across contexts. In China, the widespread acceptance of WeChat's \textit{Hongbao} feature illustrates how digital platforms can successfully adapt traditional gifting rituals without eroding their symbolic resonance. In contrast, Japanese participants expressed discomfort with digital gifting precisely because it bypassed established presentation norms such as wrapping, formality, and handwritten notes. 

These patterns suggest that culturally sensitive gifting design must go beyond surface aesthetics to embed symbolic and ritual forms into interaction design. This could include wrapping animations, ceremonial delivery flows, or regionally meaningful themes and metaphors. By honoring local gifting traditions and aesthetics, platforms can help digital gifts carry the comparable emotional and cultural weight as their physical counterparts.

\subsubsection{Design for Social Boundaries and Reciprocity as Core Gifting Functions}
Gift-giving is never a neutral act---it creates, reinforces, or tests social ties, often invoking expectations of reciprocity, relational appropriateness, and boundary management. These expectations are culturally structured and context-dependent, but they are integral to how gifting operates in any society. When digital platforms obscure or overexpose these dynamics, they risk distorting the social function of gifting--intensifying pressure, diminishing meaning, or deterring participation altogether.

Our findings reveal how platform features can both facilitate and complicate these dynamics. Particularly in Korea, visible gift histories and automated birthday reminders amplified implicit yet strong norms of reciprocal exchange, creating emotional strain in less intimate relationships. Japanese participants described limiting digital gifting to low-risk contexts, fearing misinterpretation or relational awkwardness. In contrast, Chinese participants embraced \textit{Hongbao} as a playful, socially sanctioned way to engage in gifting without triggering formal obligations.

These findings point to the need for practical design features that help users manage social expectations around reciprocity and relational boundaries. For example, in contexts where gifting may feel obligatory, users could benefit from options to hide gift prices, turn off automated reminders, or limit visibility of their gifting activity. As such, platforms should equip users with tools to adapt gifting to the relational norms of their specific social context.

\subsection{Limitation and Future Works}
While this study contributes to a deeper understanding of how culturally rooted norms shape digital gift-giving practices on messenger platforms in South Korea, Japan, and China, it has several limitations that may affect the generalizability of its findings.

First, the focus on university students and specific messenger-based gifting platforms (KakaoTalk, LINE, and WeChat) may not capture the full range of gift-giving practices across age groups or service types within each country. 


Second, while the study provides a focused cross-cultural analysis of South Korea, Japan, and China, its findings may not extend to other cultural settings. Expanding the geographic scope in future work would enable a more comprehensive understanding of global digital gifting practices.

Third, some interviews were conducted in participants' second language, which may have led to a partial loss of nuance compared to native-language interviews.

Finally, our study captures a snapshot in time. As platform features evolve and social norms shift, digital gifting practices are likely to change as well. Longitudinal studies are needed to examine how these dynamics unfold over time.

\section{Conclusion}
This study examined how digital gifting practices unfold within the cultural contexts of South Korea, Japan, and China. Through interviews with 26 university students in their 20s, we explored how traditional gift-giving norms intersect with mobile messenger-based gifting services. Our findings show that cultural values are not only reflected in how these services are used, but are also reshaped through their adoption. In Korea, digital gifting has expanded the scope of everyday relational exchanges. In Japan, it remains limited, used cautiously to preserve emotional clarity and formality. In China, the gifting of Hongbao has evolved into a playful, communal activity distinct from traditional gifting.

Beyond empirical insights, this study contributes theoretically by extending Sherry's gift-exchange model by introducing the \textit{Channel-Oriented Gifting Cycle}, which incorporates the culturally embedded decision of delivery channel selection into the planning phase. This revised model could account for how platform affordances and cultural expectations together shape the structure and meaning of digital gifts.

As a design implication, our findings highlight that digital gifting services will need to address expectations around sincerity, reciprocity, and presentation. We recommend supporting richer emotional expression, aligning design with culturally embedded rituals, and offering user control over visibility and reciprocity to accommodate diverse social norms and relational dynamics.

Taken together, this study contributes to cross-cultural HCI and CSCW by showing how digital platforms mediate and transform gifting practices, and by offering both theoretical and practical guidance for culturally sensitive design in digitally mediated gift exchange.

\received{October 2024}
\received[revised]{April 2025}
\received[accepted]{August 2025}

\begin{acks}
We thank our study participants and the reviewers for their valuable feedback and contributions. This work was supported by the SNU-Global Excellence Research
Center establishment project and the New Faculty Startup Fund (Grant No. 200-20230022) from Seoul National University.
\end{acks}

\bibliographystyle{ACM-Reference-Format}
\bibliography{gift}


\begin{thebibliography}{70}


\ifx \showCODEN    \undefined \def \showCODEN     #1{\unskip}     \fi
\ifx \showISBNx    \undefined \def \showISBNx     #1{\unskip}     \fi
\ifx \showISBNxiii \undefined \def \showISBNxiii  #1{\unskip}     \fi
\ifx \showISSN     \undefined \def \showISSN      #1{\unskip}     \fi
\ifx \showLCCN     \undefined \def \showLCCN      #1{\unskip}     \fi
\ifx \shownote     \undefined \def \shownote      #1{#1}          \fi
\ifx \showarticletitle \undefined \def \showarticletitle #1{#1}   \fi
\ifx \showURL      \undefined \def \showURL       {\relax}        \fi
\providecommand\bibfield[2]{#2}
\providecommand\bibinfo[2]{#2}
\providecommand\natexlab[1]{#1}
\providecommand\showeprint[2][]{arXiv:#2}

\bibitem[Akrich(1992)]%
        {Akrich1992de}
\bibfield{author}{\bibinfo{person}{Madeleine Akrich}.} \bibinfo{year}{1992}\natexlab{}.
\newblock \showarticletitle{The De-scription of Technical Objects}.
\newblock \bibinfo{journal}{\emph{Shaping Technology- Building Society: Studies in Sociotechnical Change}} (\bibinfo{date}{01} \bibinfo{year}{1992}).
\newblock


\bibitem[Amazon(2024)]%
        {amazon2024}
\bibfield{author}{\bibinfo{person}{Amazon}.} \bibinfo{year}{2024}\natexlab{}.
\newblock \bibinfo{title}{Amazon Gift Recommendations}.
\newblock \bibinfo{howpublished}{\url{https://www.amazon.com/gcx/Gift-recommendations/gfhz/?currency=USD&language=en_US}}.
\newblock
\newblock
\shownote{Accessed: 2024-10-23}.


\bibitem[Augustin-Jean and Saxena(2024)]%
        {augustin2024digitalising}
\bibfield{author}{\bibinfo{person}{Louis Augustin-Jean} {and} \bibinfo{person}{Vandana Saxena}.} \bibinfo{year}{2024}\natexlab{}.
\newblock \showarticletitle{Digitalising Chinese new year red packets: Changing practices and meanings}.
\newblock \bibinfo{journal}{\emph{China Perspectives}} \bibinfo{number}{136} (\bibinfo{year}{2024}), \bibinfo{pages}{21--29}.
\newblock


\bibitem[Baal(1975)]%
        {vanbaal1975reciprocity}
\bibfield{author}{\bibinfo{person}{J.~Van Baal}.} \bibinfo{year}{1975}\natexlab{}.
\newblock \bibinfo{booktitle}{\emph{The Reciprocity and The Position of Women}}.
\newblock \bibinfo{publisher}{Koubundou}, \bibinfo{address}{Tokyo, Japan}.
\newblock
\newblock
\shownote{Translated by Masako Tanaka and Satoshi Nakagawa, Koubundou, 1980}.


\bibitem[Beff(1984)]%
        {beff1984zotou}
\bibfield{author}{\bibinfo{person}{Harumi Beff}.} \bibinfo{year}{1984}\natexlab{}.
\newblock \showarticletitle{A Study on 'Zotou' as a Cultural Concept}.
\newblock In \bibinfo{booktitle}{\emph{Gift-giving in Japanese Culture}}, \bibfield{editor}{\bibinfo{person}{Mikiharu Itou} {and} \bibinfo{person}{Yasuyuki Kurita}} (Eds.). \bibinfo{publisher}{Minerva Shobo}, \bibinfo{address}{Kyoto, Japan}.
\newblock


\bibitem[Belk(1979)]%
        {belk1979gift}
\bibfield{author}{\bibinfo{person}{Russell~W Belk}.} \bibinfo{year}{1979}\natexlab{}.
\newblock \bibinfo{title}{Gift-giving behavior, Research in marketing, 2, {\'e}d. J. Sheth, Greenwich, CT}.
\newblock


\bibitem[Belk and Coon(1993)]%
        {belk1993gift}
\bibfield{author}{\bibinfo{person}{Russell~W Belk} {and} \bibinfo{person}{Gregory~S Coon}.} \bibinfo{year}{1993}\natexlab{}.
\newblock \showarticletitle{Gift giving as agapic love: An alternative to the exchange paradigm based on dating experiences}.
\newblock \bibinfo{journal}{\emph{Journal of consumer research}} \bibinfo{volume}{20}, \bibinfo{number}{3} (\bibinfo{year}{1993}), \bibinfo{pages}{393--417}.
\newblock


\bibitem[Berking(1999)]%
        {berking1999sociology}
\bibfield{author}{\bibinfo{person}{Helmuth Berking}.} \bibinfo{year}{1999}\natexlab{}.
\newblock \showarticletitle{Sociology of giving}.
\newblock  (\bibinfo{year}{1999}).
\newblock


\bibitem[Bian(2019)]%
        {bian2019guanxi}
\bibfield{author}{\bibinfo{person}{Yanjie Bian}.} \bibinfo{year}{2019}\natexlab{}.
\newblock \bibinfo{booktitle}{\emph{Guanxi, how China works}}.
\newblock \bibinfo{publisher}{John Wiley \& Sons}.
\newblock


\bibitem[Burke et~al\mbox{.}(2016)]%
        {burke2016moving}
\bibfield{author}{\bibinfo{person}{Tricia~J Burke}, \bibinfo{person}{Erin~K Ruppel}, {and} \bibinfo{person}{Dana~R Dinsmore}.} \bibinfo{year}{2016}\natexlab{}.
\newblock \showarticletitle{Moving away and reaching out: Young adults’ relational maintenance and psychosocial well-being during the transition to college}.
\newblock \bibinfo{journal}{\emph{Journal of Family Communication}} \bibinfo{volume}{16}, \bibinfo{number}{2} (\bibinfo{year}{2016}), \bibinfo{pages}{180--187}.
\newblock


\bibitem[Cadsby et~al\mbox{.}(2008)]%
        {cadsby2008trust}
\bibfield{author}{\bibinfo{person}{C~Bram Cadsby}, \bibinfo{person}{Fei Song}, {and} \bibinfo{person}{Yunyun Bi}.} \bibinfo{year}{2008}\natexlab{}.
\newblock \showarticletitle{Trust, reciprocity and social distance in China: An experimental investigation}.
\newblock \bibinfo{journal}{\emph{University of Guelph}} (\bibinfo{year}{2008}).
\newblock


\bibitem[Chen et~al\mbox{.}(2013)]%
        {chen2013chinese}
\bibfield{author}{\bibinfo{person}{Chao~C Chen}, \bibinfo{person}{Xiao-Ping Chen}, {and} \bibinfo{person}{Shengsheng Huang}.} \bibinfo{year}{2013}\natexlab{}.
\newblock \showarticletitle{Chinese Guanxi: An Integrative Review and NewDirections for Future Research}.
\newblock \bibinfo{journal}{\emph{Management and organization review}} \bibinfo{volume}{9}, \bibinfo{number}{1} (\bibinfo{year}{2013}), \bibinfo{pages}{167--207}.
\newblock


\bibitem[Choi(2024)]%
        {choi2024kakaotalk}
\bibfield{author}{\bibinfo{person}{Jeonghee Choi}.} \bibinfo{year}{2024}\natexlab{}.
\newblock \showarticletitle{Korean National Messenger 'KakaoTalk' Faces Instability... Generation Z Moving to Instagram and Telegram}.
\newblock \bibinfo{journal}{\emph{eDaily}} (\bibinfo{year}{2024}).
\newblock
\urldef\tempurl%
\url{https://m.edaily.co.kr/News/Read?newsId=03145526638983056&mediaCodeNo=257}
\showURL{%
\tempurl}
\newblock
\shownote{Accessed: 2024-08-01}.


\bibitem[Close and Kukar-Kinney(2010)]%
        {close2010beyond}
\bibfield{author}{\bibinfo{person}{Angeline~G Close} {and} \bibinfo{person}{Monika Kukar-Kinney}.} \bibinfo{year}{2010}\natexlab{}.
\newblock \showarticletitle{Beyond buying: Motivations behind consumers' online shopping cart use}.
\newblock \bibinfo{journal}{\emph{Journal of Business Research}} \bibinfo{volume}{63}, \bibinfo{number}{9-10} (\bibinfo{year}{2010}), \bibinfo{pages}{986--992}.
\newblock


\bibitem[Corp.(2010)]%
        {kakaogift2010}
\bibfield{author}{\bibinfo{person}{Kakao Corp.}} \bibinfo{year}{2010}\natexlab{}.
\newblock \bibinfo{booktitle}{\emph{The Beginning of Gifting on KakaoTalk: Launched in December 2010}}.
\newblock
\urldef\tempurl%
\url{https://www.kakaocorp.com/page/detail/5216}
\showURL{%
\tempurl}
\newblock
\shownote{Accessed on April 14, 2025}.


\bibitem[Corporation(2021)]%
        {line2021gift}
\bibfield{author}{\bibinfo{person}{LINE Corporation}.} \bibinfo{year}{2021}\natexlab{}.
\newblock \bibinfo{booktitle}{\emph{LINE GIFT Exceeds 15 Million Users}}.
\newblock
\urldef\tempurl%
\url{https://www.linecorp.com/en/pr/news/en/2021/3787}
\showURL{%
\tempurl}
\newblock
\shownote{Press release from LINE App}.


\bibitem[De~Schepper et~al\mbox{.}(2023)]%
        {de2023social}
\bibfield{author}{\bibinfo{person}{Ayla De~Schepper}, \bibinfo{person}{Noel Clycq}, {and} \bibinfo{person}{Eva Kyndt}.} \bibinfo{year}{2023}\natexlab{}.
\newblock \showarticletitle{Social networks in the transition from higher education to work: A systematic review}.
\newblock \bibinfo{journal}{\emph{Educational Research Review}}  \bibinfo{volume}{40} (\bibinfo{year}{2023}), \bibinfo{pages}{100551}.
\newblock


\bibitem[Gray et~al\mbox{.}(2014)]%
        {gray2014paying}
\bibfield{author}{\bibinfo{person}{Kurt Gray}, \bibinfo{person}{Adrian~F. Ward}, {and} \bibinfo{person}{Michael~I. Norton}.} \bibinfo{year}{2014}\natexlab{}.
\newblock \showarticletitle{Paying it Forward: Generalized Reciprocity and the Limits of Generosity}.
\newblock \bibinfo{journal}{\emph{Journal of Experimental Psychology: General}} \bibinfo{volume}{143}, \bibinfo{number}{1} (\bibinfo{year}{2014}), \bibinfo{pages}{247--254}.
\newblock
\href{https://doi.org/10.1037/a0031047}{doi:\nolinkurl{10.1037/a0031047}}


\bibitem[Gunasti and Baskin(2018)]%
        {gunasti2018200}
\bibfield{author}{\bibinfo{person}{Kunter Gunasti} {and} \bibinfo{person}{Ernest Baskin}.} \bibinfo{year}{2018}\natexlab{}.
\newblock \showarticletitle{Is a $200 Nordstrom gift card worth more or less than a $200 GAP gift card? The asymmetric valuations of luxury gift cards}.
\newblock \bibinfo{journal}{\emph{Journal of Retailing}} \bibinfo{volume}{94}, \bibinfo{number}{4} (\bibinfo{year}{2018}), \bibinfo{pages}{380--392}.
\newblock


\bibitem[Hampton et~al\mbox{.}(2011)]%
        {hampton2011social}
\bibfield{author}{\bibinfo{person}{Keith~N Hampton}, \bibinfo{person}{Lauren~Sessions Goulet}, \bibinfo{person}{Lee Rainie}, {and} \bibinfo{person}{Kristen Purcell}.} \bibinfo{year}{2011}\natexlab{}.
\newblock \bibinfo{booktitle}{\emph{Social networking sites and our lives}}. Vol.~\bibinfo{volume}{1}.
\newblock \bibinfo{publisher}{Pew Internet \& American Life Project Washington, DC}.
\newblock


\bibitem[Ho(1976)]%
        {ho1976concept}
\bibfield{author}{\bibinfo{person}{David Yau-fai Ho}.} \bibinfo{year}{1976}\natexlab{}.
\newblock \showarticletitle{On the concept of face}.
\newblock \bibinfo{journal}{\emph{American journal of sociology}} \bibinfo{volume}{81}, \bibinfo{number}{4} (\bibinfo{year}{1976}), \bibinfo{pages}{867--884}.
\newblock


\bibitem[Hui(1988)]%
        {hui1988measurement}
\bibfield{author}{\bibinfo{person}{C~Harry Hui}.} \bibinfo{year}{1988}\natexlab{}.
\newblock \showarticletitle{Measurement of individualism-collectivism}.
\newblock \bibinfo{journal}{\emph{Journal of research in personality}} \bibinfo{volume}{22}, \bibinfo{number}{1} (\bibinfo{year}{1988}), \bibinfo{pages}{17--36}.
\newblock


\bibitem[Hui and Triandis(1986)]%
        {hui1986individualism}
\bibfield{author}{\bibinfo{person}{C~Harry Hui} {and} \bibinfo{person}{Harry~C Triandis}.} \bibinfo{year}{1986}\natexlab{}.
\newblock \showarticletitle{Individualism-collectivism: A study of cross-cultural researchers}.
\newblock \bibinfo{journal}{\emph{Journal of cross-cultural psychology}} \bibinfo{volume}{17}, \bibinfo{number}{2} (\bibinfo{year}{1986}), \bibinfo{pages}{225--248}.
\newblock


\bibitem[ji~Lee(2017)]%
        {Lee2017Gifticon}
\bibfield{author}{\bibinfo{person}{Eun ji Lee}.} \bibinfo{year}{2017}\natexlab{}.
\newblock \showarticletitle{Gift Satisfaction Depending on Role and Mutual Intimacy in Gifticon-Giving Situations}.
\newblock \bibinfo{journal}{\emph{Journal of the Korea Society for Emotion and Sensibility}} \bibinfo{volume}{20}, \bibinfo{number}{3} (\bibinfo{year}{2017}), \bibinfo{pages}{131--140}.
\newblock


\bibitem[Junco and Cole-Avent(2008)]%
        {junco2008introduction}
\bibfield{author}{\bibinfo{person}{Reynol Junco} {and} \bibinfo{person}{Gail~A Cole-Avent}.} \bibinfo{year}{2008}\natexlab{}.
\newblock \showarticletitle{An introduction to technologies commonly used by college students}.
\newblock \bibinfo{journal}{\emph{New Directions for Student Services}} \bibinfo{volume}{2008}, \bibinfo{number}{124} (\bibinfo{year}{2008}), \bibinfo{pages}{3--17}.
\newblock


\bibitem[jung Kwon and young Yeo(2021)]%
        {Kwon2021MobileReturnGifts}
\bibfield{author}{\bibinfo{person}{Hye jung Kwon} {and} \bibinfo{person}{Min young Yeo}.} \bibinfo{year}{2021}\natexlab{}.
\newblock \showarticletitle{A Study on the Behavior of Recipients and Givers of Mobile Return Gifts: Intention and Expectation of Return Gifts}.
\newblock \bibinfo{journal}{\emph{Journal of Product Research}} \bibinfo{volume}{39}, \bibinfo{number}{4} (\bibinfo{year}{2021}), \bibinfo{pages}{107--115}.
\newblock


\bibitem[KIM(2024)]%
        {kim2024politics}
\bibfield{author}{\bibinfo{person}{Dong~No KIM}.} \bibinfo{year}{2024}\natexlab{}.
\newblock \showarticletitle{The Politics of Gift-Giving and Diplomatic Gifts in Traditional Korea.}
\newblock \bibinfo{journal}{\emph{Korea Journal}} \bibinfo{volume}{64}, \bibinfo{number}{2} (\bibinfo{year}{2024}).
\newblock


\bibitem[Kim(2006)]%
        {kim2006giftgiving}
\bibfield{author}{\bibinfo{person}{Jeongju Kim}.} \bibinfo{year}{2006}\natexlab{}.
\newblock \bibinfo{booktitle}{\emph{Why Do We Give and Receive Gifts? - Gifts in Cultural Sociology}}.
\newblock \bibinfo{publisher}{Samsung Economic Research Institute}, \bibinfo{address}{Seoul, South Korea}.
\newblock


\bibitem[Kiran(2024)]%
        {kiran2024wechat}
\bibfield{author}{\bibinfo{person}{Harsha Kiran}.} \bibinfo{year}{2024}\natexlab{}.
\newblock \bibinfo{booktitle}{\emph{Market Share of Messaging Apps by Country}}.
\newblock
\urldef\tempurl%
\url{https://techjury.net/blog/market-share-of-messaging-apps-by-country/}
\showURL{%
\tempurl}
\newblock
\shownote{Updated: September 10, 2024, Accessed: 2024-10-30}.


\bibitem[Kizilcec et~al\mbox{.}(2018)]%
        {kizilcec2018social}
\bibfield{author}{\bibinfo{person}{Ren{\'e}~F Kizilcec}, \bibinfo{person}{Eytan Bakshy}, \bibinfo{person}{Dean Eckles}, {and} \bibinfo{person}{Moira Burke}.} \bibinfo{year}{2018}\natexlab{}.
\newblock \showarticletitle{Social influence and reciprocity in online gift giving}. In \bibinfo{booktitle}{\emph{Proceedings of the 2018 CHI conference on human factors in computing systems}}. \bibinfo{pages}{1--11}.
\newblock


\bibitem[Koleva et~al\mbox{.}(2020)]%
        {koleva2020designing}
\bibfield{author}{\bibinfo{person}{Boriana Koleva}, \bibinfo{person}{Jocelyn Spence}, \bibinfo{person}{Steve Benford}, \bibinfo{person}{Hyosun Kwon}, \bibinfo{person}{Holger Schn{\"a}delbach}, \bibinfo{person}{Emily Thorn}, \bibinfo{person}{William Preston}, \bibinfo{person}{Adrian Hazzard}, \bibinfo{person}{Chris Greenhalgh}, \bibinfo{person}{Matt Adams}, {et~al\mbox{.}}} \bibinfo{year}{2020}\natexlab{}.
\newblock \showarticletitle{Designing hybrid gifts}.
\newblock \bibinfo{journal}{\emph{ACM Transactions on Computer-Human Interaction (TOCHI)}} \bibinfo{volume}{27}, \bibinfo{number}{5} (\bibinfo{year}{2020}), \bibinfo{pages}{1--33}.
\newblock


\bibitem[Kwon(2022)]%
        {kwon2022exploring}
\bibfield{author}{\bibinfo{person}{Hyosun Kwon}.} \bibinfo{year}{2022}\natexlab{}.
\newblock \showarticletitle{Exploring Digital Gifting Rituals}.
\newblock \bibinfo{journal}{\emph{Archives of Design Research}} \bibinfo{volume}{35}, \bibinfo{number}{2} (\bibinfo{year}{2022}), \bibinfo{pages}{73--85}.
\newblock


\bibitem[Kwon et~al\mbox{.}(2017)]%
        {kwon2017s}
\bibfield{author}{\bibinfo{person}{Hyosun Kwon}, \bibinfo{person}{Boriana Koleva}, \bibinfo{person}{Holger Schn{\"a}delbach}, {and} \bibinfo{person}{Steve Benford}.} \bibinfo{year}{2017}\natexlab{}.
\newblock \showarticletitle{" It's Not Yet A Gift" Understanding Digital Gifting}. In \bibinfo{booktitle}{\emph{Proceedings of the 2017 ACM conference on computer supported cooperative work and social computing}}. \bibinfo{pages}{2372--2384}.
\newblock


\bibitem[Lee(1988)]%
        {lee1988cross}
\bibfield{author}{\bibinfo{person}{C Lee}.} \bibinfo{year}{1988}\natexlab{}.
\newblock \showarticletitle{Cross-cultural validity of the Fishbein's behavioral intention model: Culture-bound of culture-free}.
\newblock \bibinfo{journal}{\emph{Unpublished Doctoral Dissertation, University of Texas, Austin}} (\bibinfo{year}{1988}).
\newblock


\bibitem[Lee and Green(1991)]%
        {lee1991cross}
\bibfield{author}{\bibinfo{person}{Chol Lee} {and} \bibinfo{person}{Robert~T Green}.} \bibinfo{year}{1991}\natexlab{}.
\newblock \showarticletitle{Cross-cultural examination of the Fishbein behavioral intentions model}.
\newblock \bibinfo{journal}{\emph{Journal of international business studies}}  \bibinfo{volume}{22} (\bibinfo{year}{1991}), \bibinfo{pages}{289--305}.
\newblock


\bibitem[Lee(2012)]%
        {lee2012korean}
\bibfield{author}{\bibinfo{person}{Choong~Y Lee}.} \bibinfo{year}{2012}\natexlab{}.
\newblock \showarticletitle{Korean culture and its influence on business practice in South Korea}.
\newblock \bibinfo{journal}{\emph{The Journal of International Management Studies}} \bibinfo{volume}{7}, \bibinfo{number}{2} (\bibinfo{year}{2012}), \bibinfo{pages}{184--191}.
\newblock


\bibitem[Lee et~al\mbox{.}(2020)]%
        {lee2020makes}
\bibfield{author}{\bibinfo{person}{So-Hyun Lee}, \bibinfo{person}{Su-Jin Choi}, {and} \bibinfo{person}{Hee-Woong Kim}.} \bibinfo{year}{2020}\natexlab{}.
\newblock \showarticletitle{What makes people send gifts via social network services? A mixed methods approach}.
\newblock \bibinfo{journal}{\emph{Internet Research}} \bibinfo{volume}{30}, \bibinfo{number}{1} (\bibinfo{year}{2020}), \bibinfo{pages}{315--334}.
\newblock


\bibitem[L{\'e}vi-Strauss(1969)]%
        {levi-strauss1969kinship}
\bibfield{author}{\bibinfo{person}{Claude L{\'e}vi-Strauss}.} \bibinfo{year}{1969}\natexlab{}.
\newblock \bibinfo{booktitle}{\emph{The Elementary Structures of Kinship}}.
\newblock \bibinfo{publisher}{Beacon Press}, \bibinfo{address}{Boston, MA}.
\newblock
\newblock
\shownote{Originally published in 1949 in French}.


\bibitem[Li et~al\mbox{.}(2022)]%
        {li2022differentiating}
\bibfield{author}{\bibinfo{person}{Peikai Li}, \bibinfo{person}{Jian-Min Sun}, {and} \bibinfo{person}{Toon~W Taris}.} \bibinfo{year}{2022}\natexlab{}.
\newblock \showarticletitle{Differentiating between gift giving and bribing in China: A guanxi perspective}.
\newblock \bibinfo{journal}{\emph{Ethics \& Behavior}} \bibinfo{volume}{32}, \bibinfo{number}{4} (\bibinfo{year}{2022}), \bibinfo{pages}{307--325}.
\newblock


\bibitem[Liu(2023)]%
        {liu2023wechat}
\bibfield{author}{\bibinfo{person}{Ruoxi Liu}.} \bibinfo{year}{2023}\natexlab{}.
\newblock \showarticletitle{WeChat online visual language among Chinese Gen Z: virtual gift, aesthetic identity, and affection language}.
\newblock \bibinfo{journal}{\emph{Frontiers in Communication}}  \bibinfo{volume}{8} (\bibinfo{year}{2023}), \bibinfo{pages}{1172115}.
\newblock


\bibitem[Lonner et~al\mbox{.}(1980)]%
        {lonner1980culture}
\bibfield{author}{\bibinfo{person}{Walter~J Lonner}, \bibinfo{person}{John~W Berry}, {and} \bibinfo{person}{Geert~H Hofstede}.} \bibinfo{year}{1980}\natexlab{}.
\newblock \showarticletitle{Culture's consequences: International differences in work-related values}.
\newblock \bibinfo{journal}{\emph{University of Illinois at Urbana-Champaign's Academy for Entrepreneurial Leadership Historical Research Reference in Entrepreneurship}} (\bibinfo{year}{1980}).
\newblock


\bibitem[Luo(1997)]%
        {luo1997guanxi}
\bibfield{author}{\bibinfo{person}{Yadong Luo}.} \bibinfo{year}{1997}\natexlab{}.
\newblock \showarticletitle{Guanxi: Principles, philosophies, and implications}.
\newblock \bibinfo{journal}{\emph{Human systems management}}  \bibinfo{volume}{16} (\bibinfo{year}{1997}), \bibinfo{pages}{43--52}.
\newblock


\bibitem[Malinowski(1978)]%
        {malinowski1978argonauts}
\bibfield{author}{\bibinfo{person}{Bronislaw Malinowski}.} \bibinfo{year}{1978}\natexlab{}.
\newblock \bibinfo{booktitle}{\emph{Argonauts of the Western Pacific: An Account of Native Enterprise and Adventure in the Archipelagoes of Melanesian New Guinea}}.
\newblock \bibinfo{publisher}{Routledge}.
\newblock


\bibitem[Mauss(1954)]%
        {mauss1954gift}
\bibfield{author}{\bibinfo{person}{Marcel Mauss}.} \bibinfo{year}{1954}\natexlab{}.
\newblock \bibinfo{booktitle}{\emph{The Gift: Forms and Functions of Exchange in Archaic Societies}}.
\newblock \bibinfo{publisher}{W.W. Norton \& Company}.
\newblock


\bibitem[Mauss(1967)]%
        {mauss1967gift}
\bibfield{author}{\bibinfo{person}{Marcel Mauss}.} \bibinfo{year}{1967}\natexlab{}.
\newblock \showarticletitle{The gift: forms and functions ofexchange in Archaic societies}.
\newblock \bibinfo{journal}{\emph{Trans. Ian Cunnison (New York: WW Norton and Co., 1967)}} (\bibinfo{year}{1967}), \bibinfo{pages}{80}.
\newblock


\bibitem[News(2024)]%
        {wechat2024Hongbao2}
\bibfield{author}{\bibinfo{person}{Tencent News}.} \bibinfo{year}{2024}\natexlab{}.
\newblock \bibinfo{booktitle}{\emph{WeChat Red Envelope 2.0: Speculations After Its Launch}}.
\newblock
\urldef\tempurl%
\url{https://news.qq.com/rain/a/20241223A064F600?utm_source=chatgpt.com}
\showURL{%
\tempurl}
\newblock
\shownote{Published in Beijing by a Technology Creator on Tencent News}.


\bibitem[Niu et~al\mbox{.}(2022)]%
        {niu2022machiavellianism}
\bibfield{author}{\bibinfo{person}{Gengfeng Niu}, \bibinfo{person}{Xiaohan Shi}, \bibinfo{person}{Siyu Jin}, \bibinfo{person}{Wencheng Yang}, \bibinfo{person}{Yang Wu}, {and} \bibinfo{person}{Xiaojun Sun}.} \bibinfo{year}{2022}\natexlab{}.
\newblock \showarticletitle{Machiavellianism and gift-giving in live video streaming: The mediating role of desire for control and the moderating role of materialism}.
\newblock \bibinfo{journal}{\emph{Behavioral Sciences}} \bibinfo{volume}{12}, \bibinfo{number}{5} (\bibinfo{year}{2022}), \bibinfo{pages}{157}.
\newblock


\bibitem[Orlikowski(1992)]%
        {orlikowski1992duality}
\bibfield{author}{\bibinfo{person}{Wanda~J Orlikowski}.} \bibinfo{year}{1992}\natexlab{}.
\newblock \showarticletitle{The duality of technology: Rethinking the concept of technology in organizations}.
\newblock \bibinfo{journal}{\emph{Organization science}} \bibinfo{volume}{3}, \bibinfo{number}{3} (\bibinfo{year}{1992}), \bibinfo{pages}{398--427}.
\newblock


\bibitem[Park(1998)]%
        {park1998comparison}
\bibfield{author}{\bibinfo{person}{Seong-Yeon Park}.} \bibinfo{year}{1998}\natexlab{}.
\newblock \showarticletitle{A comparison of Korean and American gift-giving behaviors}.
\newblock \bibinfo{journal}{\emph{Psychology \& Marketing}} \bibinfo{volume}{15}, \bibinfo{number}{6} (\bibinfo{year}{1998}), \bibinfo{pages}{577--593}.
\newblock


\bibitem[Patton(2014)]%
        {patton2014qualitative}
\bibfield{author}{\bibinfo{person}{Michael~Quinn Patton}.} \bibinfo{year}{2014}\natexlab{}.
\newblock \bibinfo{booktitle}{\emph{Qualitative research \& evaluation methods: Integrating theory and practice}}.
\newblock \bibinfo{publisher}{Sage publications}.
\newblock


\bibitem[Rupp(2003)]%
        {rupp2003gift}
\bibfield{author}{\bibinfo{person}{Katherine Rupp}.} \bibinfo{year}{2003}\natexlab{}.
\newblock \bibinfo{booktitle}{\emph{Gift-giving in Japan: Cash, connections, cosmologies}}.
\newblock \bibinfo{publisher}{Stanford University Press}.
\newblock


\bibitem[Sandel(1998)]%
        {sandel1998money}
\bibfield{author}{\bibinfo{person}{Michael~J Sandel}.} \bibinfo{year}{1998}\natexlab{}.
\newblock \bibinfo{booktitle}{\emph{What money can't buy: the moral limits of markets}}.
\newblock \bibinfo{publisher}{Brasenose College, Oxford}.
\newblock


\bibitem[Saxena(2024)]%
        {saxena2024digitalising}
\bibfield{author}{\bibinfo{person}{Vandana Saxena}.} \bibinfo{year}{2024}\natexlab{}.
\newblock \showarticletitle{Digitalising Chinese new year red packets: Changing practices and meanings}.
\newblock \bibinfo{journal}{\emph{China Perspectives}} \bibinfo{number}{136} (\bibinfo{year}{2024}), \bibinfo{pages}{21--29}.
\newblock


\bibitem[Schwartz(1967)]%
        {schwartz1967social}
\bibfield{author}{\bibinfo{person}{Barry Schwartz}.} \bibinfo{year}{1967}\natexlab{}.
\newblock \showarticletitle{The social psychology of the gift}.
\newblock \bibinfo{journal}{\emph{American journal of Sociology}} \bibinfo{volume}{73}, \bibinfo{number}{1} (\bibinfo{year}{1967}), \bibinfo{pages}{1--11}.
\newblock


\bibitem[seok Lee(2013)]%
        {Lee2013DigitalGifts}
\bibfield{author}{\bibinfo{person}{Han seok Lee}.} \bibinfo{year}{2013}\natexlab{}.
\newblock \showarticletitle{A Qualitative Study on Digital Content Gift-Giving Behavior in Online Environments}.
\newblock \bibinfo{journal}{\emph{Cyber Society and Culture}} \bibinfo{volume}{4}, \bibinfo{number}{1} (\bibinfo{year}{2013}), \bibinfo{pages}{93--113}.
\newblock


\bibitem[Sherry~Jr(1983)]%
        {sherry1983gift}
\bibfield{author}{\bibinfo{person}{John~F Sherry~Jr}.} \bibinfo{year}{1983}\natexlab{}.
\newblock \showarticletitle{Gift giving in anthropological perspective}.
\newblock \bibinfo{journal}{\emph{Journal of consumer research}} \bibinfo{volume}{10}, \bibinfo{number}{2} (\bibinfo{year}{1983}), \bibinfo{pages}{157--168}.
\newblock


\bibitem[Silverstone and Haddon(1996)]%
        {silverstone1996design}
\bibfield{author}{\bibinfo{person}{Roger Silverstone} {and} \bibinfo{person}{Leslie Haddon}.} \bibinfo{year}{1996}\natexlab{}.
\newblock \showarticletitle{Design and the domestication of information and communication technologies: Technical change and everyday life}.
\newblock  (\bibinfo{year}{1996}).
\newblock


\bibitem[Spence et~al\mbox{.}(2023)]%
        {spence2023more}
\bibfield{author}{\bibinfo{person}{Jocelyn Spence}, \bibinfo{person}{Boriana Koleva}, \bibinfo{person}{Steve Benford}, \bibinfo{person}{Dimitrios Darzentas}, \bibinfo{person}{Martin Flintham}, \bibinfo{person}{Kevin Glover}, \bibinfo{person}{Hanne Wagner}, \bibinfo{person}{Rebecca Gibson}, {and} \bibinfo{person}{Emily-Clare Thorn}.} \bibinfo{year}{2023}\natexlab{}.
\newblock \showarticletitle{"More than a cliche": Experiencing Hybrid Gifting in the Wild}.
\newblock \bibinfo{journal}{\emph{ACM Transactions on Computer-Human Interaction}} \bibinfo{volume}{30}, \bibinfo{number}{4} (\bibinfo{year}{2023}), \bibinfo{pages}{1--31}.
\newblock


\bibitem[{Straight Team}(2024)]%
        {straight2024line}
\bibfield{author}{\bibinfo{person}{{Straight Team}}.} \bibinfo{year}{2024}\natexlab{}.
\newblock \showarticletitle{Why Does Japan Target LINE? - The Mystery of 'Re-evaluating Capital Relationships'}.
\newblock \bibinfo{journal}{\emph{MBC News}} (\bibinfo{year}{2024}).
\newblock
\urldef\tempurl%
\url{https://imnews.imbc.com/replay/straight/6604117_28993.html}
\showURL{%
\tempurl}
\newblock
\shownote{Accessed: 2024-06-02}.


\bibitem[Suarez et~al\mbox{.}(2020)]%
        {suarez2020understanding}
\bibfield{author}{\bibinfo{person}{Lina Suarez}, \bibinfo{person}{Nichole Hugo}, {and} \bibinfo{person}{C Paris}.} \bibinfo{year}{2020}\natexlab{}.
\newblock \showarticletitle{Understanding Japanese consumer behaviour and cultural relevance of gift giving}.
\newblock \bibinfo{journal}{\emph{African Journal of Hospitality, Tourism and Leisure}} \bibinfo{volume}{8}, \bibinfo{number}{4} (\bibinfo{year}{2020}).
\newblock


\bibitem[Suzuki(1988)]%
        {suzuki1988gift}
\bibfield{author}{\bibinfo{person}{Takemitsu Suzuki}.} \bibinfo{year}{1988}\natexlab{}.
\newblock \showarticletitle{Gift-Giving and Social Relationships}.
\newblock \bibinfo{journal}{\emph{Annual Review of Sociological Studies}} \bibinfo{volume}{1988}, \bibinfo{number}{1} (\bibinfo{year}{1988}), \bibinfo{pages}{23--34}.
\newblock
\href{https://doi.org/10.5690/kantoh.1988.23}{doi:\nolinkurl{10.5690/kantoh.1988.23}}


\bibitem[Taylor and Harper(2002)]%
        {taylor2002age}
\bibfield{author}{\bibinfo{person}{Alex~S Taylor} {and} \bibinfo{person}{Richard Harper}.} \bibinfo{year}{2002}\natexlab{}.
\newblock \showarticletitle{Age-old practices in the'new world' a study of gift-giving between teenage mobile phone users}. In \bibinfo{booktitle}{\emph{Proceedings of the SIGCHI conference on Human factors in computing systems}}. \bibinfo{pages}{439--446}.
\newblock


\bibitem[Triandis(1990)]%
        {triandis1990cross}
\bibfield{author}{\bibinfo{person}{Harry~C Triandis}.} \bibinfo{year}{1990}\natexlab{}.
\newblock \showarticletitle{Cross-cultural studies of individualism and collectivism.}
\newblock  (\bibinfo{year}{1990}).
\newblock


\bibitem[Ulkuniemi(2022)]%
        {ulkuniemi2022comparison}
\bibfield{author}{\bibinfo{person}{Nora Ulkuniemi}.} \bibinfo{year}{2022}\natexlab{}.
\newblock \bibinfo{title}{Comparison of Japanese and Finnish Gift Giving Behavior}.
\newblock


\bibitem[Walder(1988)]%
        {walder1988communist}
\bibfield{author}{\bibinfo{person}{Andrew~G Walder}.} \bibinfo{year}{1988}\natexlab{}.
\newblock \bibinfo{booktitle}{\emph{Communist neo-traditionalism: Work and authority in Chinese industry}}.
\newblock \bibinfo{publisher}{Univ of California Press}.
\newblock


\bibitem[Watts and Shmargad(2015)]%
        {watts2015social}
\bibfield{author}{\bibinfo{person}{Jameson~KM Watts} {and} \bibinfo{person}{Yotam Shmargad}.} \bibinfo{year}{2015}\natexlab{}.
\newblock \showarticletitle{Social visibility and the gifting of digital goods}. In \bibinfo{booktitle}{\emph{Proceedings of the 2015 ACM on Conference on Online Social Networks}}. \bibinfo{pages}{49--58}.
\newblock


\bibitem[Weinberger et~al\mbox{.}(2025)]%
        {weinberger2025relational}
\bibfield{author}{\bibinfo{person}{Michelle~F Weinberger}, \bibinfo{person}{Ernest Baskin}, {and} \bibinfo{person}{Kunter Gunasti}.} \bibinfo{year}{2025}\natexlab{}.
\newblock \showarticletitle{Relational gifting: Conceptual frameworks and an agenda for a new generation of research}.
\newblock \bibinfo{journal}{\emph{Journal of Consumer Research}} \bibinfo{volume}{51}, \bibinfo{number}{6} (\bibinfo{year}{2025}), \bibinfo{pages}{1252--1278}.
\newblock


\bibitem[Wu and Ma(2017)]%
        {wu2017money}
\bibfield{author}{\bibinfo{person}{Ziming Wu} {and} \bibinfo{person}{Xiaojuan Ma}.} \bibinfo{year}{2017}\natexlab{}.
\newblock \showarticletitle{Money as a social currency to manage group dynamics: Red packet gifting in chinese online communities}. In \bibinfo{booktitle}{\emph{Proceedings of the 2017 CHI Conference Extended Abstracts on Human Factors in Computing Systems}}. \bibinfo{pages}{2240--2247}.
\newblock


\bibitem[Xin and Pearce(1996)]%
        {xin1996guanxi}
\bibfield{author}{\bibinfo{person}{Katherine~K Xin} {and} \bibinfo{person}{Jone~L Pearce}.} \bibinfo{year}{1996}\natexlab{}.
\newblock \showarticletitle{Guanxi: Connections as substitutes for formal institutional support}.
\newblock \bibinfo{journal}{\emph{Academy of management journal}} \bibinfo{volume}{39}, \bibinfo{number}{6} (\bibinfo{year}{1996}), \bibinfo{pages}{1641--1658}.
\newblock


\bibitem[Yang et~al\mbox{.}(2011)]%
        {yang2011virtual}
\bibfield{author}{\bibinfo{person}{Jiang Yang}, \bibinfo{person}{Mark~S Ackerman}, {and} \bibinfo{person}{Lada~A Adamic}.} \bibinfo{year}{2011}\natexlab{}.
\newblock \showarticletitle{Virtual gifts and guanxi: supporting social exchange in a chinese online community}. In \bibinfo{booktitle}{\emph{Proceedings of the ACM 2011 conference on Computer supported cooperative work}}. \bibinfo{pages}{45--54}.
\newblock


\end{thebibliography}
\end{CJK}
\end{document}